\documentclass[11pt,onecolumn]{article}
\usepackage{times,graphics,epsfig,amsmath,shapepar,fancyhdr,amsthm}
\usepackage{graphicx,amsfonts,amssymb,url,float,verbatim}
\usepackage[usenames]{color}
\usepackage{url}
\newtheorem{thm}{Theorem}
\newtheorem{lem}[thm]{Lemma}
\newtheorem{cor}[thm]{Corollary}
{\makeatletter
 \gdef\xxxmark{%
   \expandafter\ifx\csname @mpargs\endcsname\relax % in minipage?
     \expandafter\ifx\csname @captype\endcsname\relax % in figure/caption?
       \marginpar{xxx}% not in a caption or minipage, can use marginpar
     \else
       xxx % notice trailing space
     \fi
   \else
     xxx % notice trailing space
   \fi}
 \gdef\xxx{\@ifnextchar[\xxx@lab\xxx@nolab}
 \long\gdef\xxx@lab[#1]#2{{\bf [\xxxmark #2 ---{\sc #1}]}}
 \long\gdef\xxx@nolab#1{{\bf [\xxxmark #1]}}
 \gdef\turnoffxxx{\long\gdef\xxx@lab[##1]##2{}\long\gdef\xxx@nolab##1{}}%
}

\title{Reconfiguration of 3D Crystalline Robots Using $O(\log n)$ Parallel Moves
\thanks{
This result, constrained to 2D, 
was presented by the authors at the % ISAAC 2008
19th International Symposium on Algorithms and Computation~\cite{isaac08}.
}}

\author{Greg Aloupis\thanks
{Universit\'e Libre de Bruxelles.   aloupis.greg@gmail.com}
\and
S\'ebastien Collette\thanks
{Universit\'e Libre de Bruxelles. Charg\'e de Recherches du FRS-FNRS.
secollet@ulb.ac.be} 
\and
Erik D. Demaine\thanks
{Massachusetts Institute of Technology.  edemaine@mit.edu
Partially supported by NSF CAREER award CCF-0347776, DOE grant DE-FG02-04ER25647, and AFOSR grant FA9550-07-1-0538.}
\and
Stefan Langerman\thanks
{Universit\'e Libre de Bruxelles. Ma\^itre de Recherches du FRS-FNRS.   slanger@ulb.ac.be} \and
Vera Sacrist\'an\thanks
{Universitat Polit\`ecnica de Catalunya. vera.sacristan@upc.edu
Partially supported by projects MEC MTM2006-01267 and Gen. Cat. DGR 2009SGR1040.} 
\and
Stefanie Wuhrer\thanks
{Carleton University. stefanie.wuhrer@gmail.com}
}

\begin{document}

\maketitle

\begin{abstract}
We consider the theoretical model of Crystalline robots, which have been
introduced and prototyped by the robotics community.
These robots consist of independently manipulable unit-square atoms that can
extend/contract arms on each side and attach/detach from neighbors.  These
operations suffice to reconfigure between any two given (connected) shapes.
The worst-case number of sequential moves required to transform one
connected configuration to another
% (up to a constant-factor resolution)
is known to be $\Theta(n)$.  %at ISAAC 2007.
%But the algorithm makes moves sequentially,
However, in principle, atoms can all move simultaneously.  We develop
a parallel algorithm for reconfiguration that runs in only $O(\log n)$
parallel steps, although the total number of operations increases slightly
to $\Theta(n \log n)$.
The result is the first (theoretically) almost-instantaneous universally
reconfigurable robot built from simple units.

%Let $S$
%and $T$ be configurations of a two-dimensional lattice robot, composed of $n$ cubic atoms arranged in $k{\times}k$  meta-modules, where $k$ is constant.
%In this paper we propose novel algorithms that reconfigure $S$ into $T$.
%The reconfiguration involves a total of
%$O(n \log{n})$ atom operations (expand, contract, attach, detach) and is
%performed in $O(\log{n})$ parallel steps.

%This improves on our previous reconfiguration algorithm, which takes $O(n)$ parallel steps.

%\begin{center}\fbox{\begin{minipage}[h]{0.85\textwidth}\begin{note}
%This draft only considers $k=8$.
%\end{note}\end{minipage}}\end{center}

\end{abstract}

\section{Introduction}

\begin{figure}[htb]
\centering
\includegraphics[width = 12.3cm]{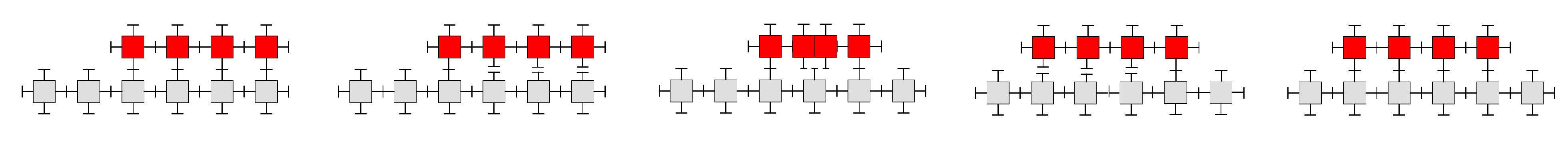}
\caption{\emph{Example of reconfiguring Crystalline atoms: the top row of atoms is able to shift to the left, using the bottom row of atoms as a fixed base.}}
\label{example}
\end{figure}

In this paper, we consider homogeneous self-reconfiguring modular robots
composed of unit-cube \emph{atoms} arranged in a grid configuration.
Each atom is equipped with mechanisms allowing it to extend each face
out one unit and later retract it back.  Furthermore, the faces can
attach/detach to faces of adjacent atoms. At all times, the atoms should
form a connected configuration.  When groups of atoms perform the four basic atom
operations (expand, contract, attach, detach) in a coordinated way, the
atoms move relative to one another, resulting in a reconfiguration of the
robot.  Fig.~\ref{example} shows an example of such a reconfiguration.
Each atom is depicted as a square, with a $T$-shaped arm on each side.

The robotics community has implemented this model in two prototype systems:
\emph{Crystalline atoms} \cite{BFR02,RV01} and
\emph{Telecube atoms} \cite{SHY02}.
%% The following is really only done in the algorithms;
%% not clear the hardware has any preference, except that all-expanded
%% is probably only structurally sound in a 2D world...
In the Crystalline model, the default state for atoms is expanded,
while in the Telecube model, the default state is contracted.
Thus Fig.~\ref{example} reconfigures a Crystalline robot,
or an expanded Telecube robot.
%Crystalline robots work in a single plane, forbidding expand / contract / attach / detach operations parallel to the $z$ axis, which is the case we consider in this paper. \textcolor{red}{Is this an old statement?}

To ensure connectedness of the configuration space, atoms must be
arranged in  \emph{modules},
which are groups of $k{\times}k{\times}k$ atoms.
Any value $k \geq 2$ suffices for universal reconfigurability \cite{cbots-cgta,VYS02}.
Here the collection of atoms composing a robot must remain \emph{connected}
in the sense that its module graph (where vertices correspond to modules
and edges correspond to attached modules) is connected.

The complexity of a reconfiguration algorithm can be
measured by the number of \emph{parallel steps} performed (``makespan''),
as well as the total number of atom operations (``work'').
In a parallel step, many atoms may perform moves simultaneously.
The number of parallel steps is typically the most significant factor in
overall reconfiguration time, because the mechanical actions
(expansion, contraction, attachment, detachment) are
the slowest part of the system.

Our main contribution in this paper is a reconfiguration algorithm that,
given a source robot $S$ and a target robot $T$, each composed of $n$ atoms
arranged in $k{\times}k{\times}k$ modules for some constant~$k$, reconfigures $S$ into
$T$ in $O(\log n)$ parallel steps and a total of $O(n \log n)$ atom
operations.
%This result improves upon the reconfiguration time of the algorithm presented by Aloupis et al.~\cite{cbots-cgta}, which takes $O(n)$ parallel steps (although only $O(n)$ total operations, and also for three-dimensional robots), as well as previous $O(n^2)$ algorithms \cite{RV01,VYS02,BR03}.

A central assumption in our algorithm is that one atom,
by contracting or expanding, can pull or push all $n$ atoms
(\emph{linear strength}).
Thus our algorithm certainly tests the structural limits of a
modular robot, but on the other hand this assumption enables us
to achieve reconfiguration times that are likely asymptotically optimal.
\begin{comment}
The quadratic reconfiguration algorithms of~\cite{RV01,VYS02,BR03}
may be given credit for being the least physically demanding
on the structure of the robot.
Even the algorithm in~\cite{cbots-cgta} is less demanding than what we propose
here, because it does not produce arbitrarily high velocities
(although it still uses linear strength).
Another recent algorithm~\cite{wafr08} considers the case where atoms have
only constant strength, and attains $O(n)$ parallel steps and
$O(n^2)$ total operations, which is optimal in this setting.
Thus the improvement in reconfiguration time obtained here \emph{requires}
a more relaxed physical model.
\end{comment}

The main idea of our parallel algorithm is to reconfigure the given robot into a canonical form, by recursively dividing the plane
into a hierarchy of cube-shaped \emph{cells} and employing a divide-and-conquer
technique to merge adjacent cells.
Each merge creates a cell containing a simple structure
using a constant number of moves.
This structure, which fills the perimeter of a cell as much as possible,
can be decomposed into a constant number of rectangular components.
Because the steps to merge cells of the same level can be executed
in parallel, the total number of parallel steps used to reconfigure
any configuration to a simple structure is $O(\log n)$.
% An additional $O(1)$ parallel steps reconfigure the final shape into a canonical form, which
The entire reconfiguration takes place in the smallest box with side $2^h$ that
contains the initial configuration, where $h$ is an integer.

We choose to describe our algorithm in terms of the naturally expanded modules
of Crystalline robots.  Of course, this immediately implies reconfigurability
in the naturally contracted Telecube model, by adding one step at the beginning
and end in which all atoms expand and contract in parallel.
We also expect that the individual constructions in our algorithm
can be modified to directly work in the Telecube model as well.

In sections~\ref{sec-def}--\ref{k=32} we focus on 2D Crystalline atoms.  In this case,
our algorithm effectively uses modules of  $4{\times}4$ atoms, but
for clarity and brevity assumes that atoms initially appear in \emph{blocks} of
 $32{\times}32$.
Reducing the module size leads to more complicated basic operations
that we have designed for use on large rectangular components.
On the other hand, reducing the initial block size leads to a larger number of
possible shapes that we must consider during the merge of cells.
We have designed (though not rigorously analyzed) a range of algorithms
for $2{\times}2$ modules with decreasing restrictions on block size.
This is discussed in our concluding remarks (section~\ref{discussion}).
However, the bulk of this paper focuses on the version
that is easiest to describe.

In section~\ref{sec-3D} we describe how to extend our algorithm from 2D to 3D.
Although the technique used is the same, there are a few subtle differences that  we
address, while attempting to avoid repetition.   The block size and/or module size
increase slightly, depending on the chosen extension approach.

% Our block size in 3D is\textcolor{red}{TBD}

\section{Related Work}

Robots designed with a unique purpose can be very efficient but
may lack versatility in the sense that they may not efficiently perform tasks
other than the ones they were initially designed for. Furthermore, problems can occur when the robots actuate in an environment
different from that originally envisaged.

For this reason, research on self-reconfiguring robots is active in the robots community. These robots can self-reconfigure in order
to adapt to unknown environments and perform unknown tasks. Modular
robots can self-organize to repair or replace damaged structures,
transport pieces of machinery or manipulate objects in inaccessible 
locations. Their adaptability allows them to climb over obstacles,
maneuver  into narrow passages, and change
shape to adapt to their environment. In addition, their modularity
enables auto-repairing when one unit fails and, of course, allows
locomotion. Two of the main goals in this field are to
construct new self-reconfiguring systems and to design efficient algorithms for their autonomous reconfiguration.    

Several models have been proposed, most of which have been
prototyped.  Crystalline robots are an example of a (cube) lattice model. Alternative models include hexagonal lattice
robots~\cite{PCSC96}, 3-dimensional rectangular grid
robots~\cite{HTFKAKE98}, chain robots such as \emph{Polybot}~\cite{YDF00}, and
 mixed (chain and lattice) models as \emph{M-Tran}~\cite{YMKTKK02}. 
 For a survey on the large range of modular robot
prototypes, as well as for an outlook on the future goals and prospects of the field,
see~\cite{survey2,survey1}.

%\emph{Crystalline} and \emph{Telecube} robots, which belong to the class of {\em lattice} modular robots, have the property that their units can contract and expand. Hence, it is possible to envisage reconfiguring algorithms for this kind of robots that do not require extra space to carry out the reconfiguration. This issue is of special relevance when working in restricted and limited environments.  Furthermore, these robots do not require the reconfiguration to be produced by translating the robot units along the surface of the robot from their initial to their goal position, unlike most previous prototypes. \textcolor{red}{I dont really understand this paragraph.  Expansion/Contraction is a vague concept, and I doubt the reader will make a connection to in-place reconfiguration or how the robot will perform locomotion. Also we should not write in this section as if }

A $O(n^2)$-time sequential algorithm for reconfiguring general classes of lattice-based robotic models was given in~\cite{CPE96}, and it was also shown that this number of individual atom moves is sometimes required.  The first reconfiguration algorithm specifically proposed for Crystalline robots was the \emph{melt-grow} algorithm~\cite{RV01}. It reconfigures one shape
to another by constructing a strip of modules as an intermediate
step. This is done in  $O(n^2)$
steps.  The algorithm is not in-place; the additional area used is $O(n)$, in terms of
the unit-volume of a module.
The \emph{PAC-MAN}
algorithm~\cite{BR03} and its variant~\cite{VYS02} are in-place, 
have a more parallelized nature, but still use $O(n^2)$ steps. In fact, for each of these algorithms there exist simple instances that require
$\Omega(n^2)$ steps.

A linear-time parallel algorithm for reconfiguring within the
bounding box of source and target is given in~\cite{cbots-cgta}. The
total number of individual moves is also linear. No
restrictions are made concerning physical properties of the robots.
For example, $O(n)$ strength is required, since modules can carry
towers and push large masses during certain operations.

However, it turns out that if the strength of each atom is restricted to $O(1)$, i.e. one atom 
can only pull or
push a fixed number of other atoms, any Crystalline  robot can still be reconfigured in-place, using $O(n)$ parallel
steps~\cite{wafr08}. Unavoidably, the total number of individual moves is
$O(n^2)$.  This is shown to be worst-case optimal. 

When $O(1)$ strength is required but velocities are allowed to grow arbitrarily over time, 
reconfiguration takes  $\Theta(\sqrt{n})$ steps~\cite{reif}.  Note that this is
for 2D robots and  the third dimension is used as an intermediate. 

In this paper, we do not restrict the strength or maximum velocity
of each robot.   In this model, we can reconfigure any
shape using $O(\log n)$ parallel steps.

%%%
%Algorithms for reconfiguring Crystalline and Telecube robots in
%$O(n^2)$ parallel steps have been given in~\cite{RV01,VYS02,BR03}.
%
%A linear-time parallel algorithm for reconfiguring within the
%bounding box of source and target is given in~\cite{cbots-cgta}. The
%total number of individual moves is also linear. However, no
%restrictions are made concerning physical properties of the robots.
%For example, $O(n)$ strength is required, since modules can carry
%towers and push large masses during certain operations.
%
%A $O(\sqrt{n})$ time algorithm for 2D robots, using the third
%dimension as an intermediate, is given in~\cite{reif}.  This is
%optimal in the model considered, which permits linear velocities,
%but only constant acceleration.
%
%If the strength of each robot is restricted to $O(1)$, a robot can
%be reconfigured using $O(n)$ parallel steps~\cite{wafr08}. The total
%number of individual moves is $O(n^2)$. This is shown to be
%worst-case optimal.
%
%In this paper, we do not restrict the strength or maximum velocity
%of each robot. We show that in this case, we can reconfigure in
%$O(\log n)$ parallel steps.
%%%

\section{Definitions}
\label{sec-def}
%A \emph{ module graph} has a node for each  module of the robot. Two nodes are connected in the  module graph if the corresponding  modules are adjacent.
%\begin{definition}
Since we will first focus on 2D reconfigurations (in sections~\ref{basic} and~\ref{k=32}), our basic definitions given here will
be for 2D as well.   In section~\ref{sec-3D} we will describe how these definitions can
be extended if necessary.

\begin{figure}[htb]
\centering
\begin{tabular}{c @{\qquad} c @{\qquad} c}
\includegraphics[height = 2.0cm]{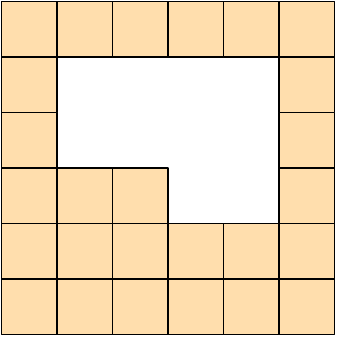} &
\includegraphics[height = 2.0cm]{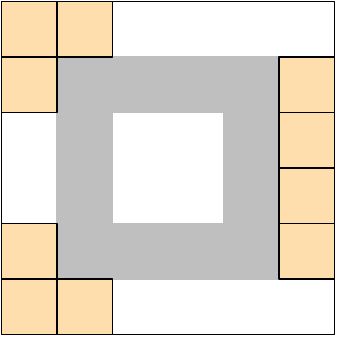} &
\includegraphics[height = 2.0cm]{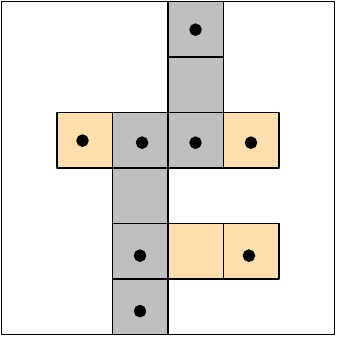} \\
(a) & (b) & (c)\\
\end{tabular}
\caption{\emph{Definitions; modules are depicted by squares. (a) A ring. (b) A sparse cell with five side-branches and shaded near-boundary. (c) A shaded backbone and eight nodes.}}
\label{fig-defs}
\end{figure}

We will mainly deal with modules, not atoms,
which can be viewed as lying on their own square lattice
somewhat coarser than the atom lattice.
Refer to Fig.~\ref{fig-defs} for examples of the following notions.
In all figures, modules are depicted as squares unless mentioned otherwise.
A module is a \emph{node} if it has exactly one neighbor
(a leaf node), more than two neighbors (a branching node),
or exactly two neighbors not collinear with the node (a bending node).
%\end{definition}
%\begin{definition}
A \emph{branch} is a straight path of (non-node) modules between two nodes
(including the nodes themselves).
%\end{definition}
%\begin{definition}
A \emph{cell} is a square of module positions (aligned with the module
lattice), some of which may be occupied by modules.
The \emph{boundary} of a cell consists of all module positions
touching the cell's border.
%Those positions are defined to have zero distance from the border, where distance is measured in  module units.
%\end{definition}
%\begin{definition}
For cells of sufficient size % ($\geq 4$ modules per side),
the \emph{near-boundary} consists of all module positions
% at distance one from the border (i.e., all positions
adjacent to the cell's boundary.
% More generally, the \emph{$k$th order near-boundary} consists of all  module positions at distance $k$ from the border.
%\end{definition}
%\begin{definition}
If a branch lies entirely in the boundary of a cell,
we call it a \emph{side-branch}.
% (resp. \emph{near-blob})
%%%%%%%if it lies entirely on the boundary
% (resp. \emph{near-boundary})
%of a cell.
%\end{definition}
%\begin{definition}
%A branch is a \emph{near-blob} if it lies entirely in the near-boundary of a cell.
% and touches only a set of blobs.
%\end{definition}
%\begin{definition}
The configuration within a cell is a \emph{ring} if the entire cell's boundary
is occupied by modules, and all remaining modules within the cell are arranged
%in a ``highly symmetric'' shape.
%Specifically, we choose to arrange inner modules
at the bottom of the cell,
filling row by row from left to right.
%\end{definition}
%\begin{definition}
The configuration within a cell is \emph{sparse}
if it contains only side-branches.
% and at most two near-blobs per side.
%\end{definition}
%\begin{definition}
A \emph{backbone} is a set of branches forming a path that connects
two opposite edges of a cell.
%\end{definition}

\section{Elementary Moves That Use $O(1)$ Parallel Steps}
\label{basic}
%Let a module consist of $4{\times}4$ atoms and $R$ be a robot configuration of $n$  modules. 
%This section describes some elementary moves
% applied to $R$     %%%%  R was left undefined
%that can be executed in $O(1)$ parallel steps.
%%%%%%%%%%%  we say that clearly in the section title %%%%

Throughout this paper, whenever we describe a move, it is implied that we do
not disconnect the robot and that no collisions occur.

\begin{figure}[htb]
\centering
%\begin{tabular}{c c}
\begin{tabular}{c @{\qquad\qquad} c}
\includegraphics[height = 3.5cm]{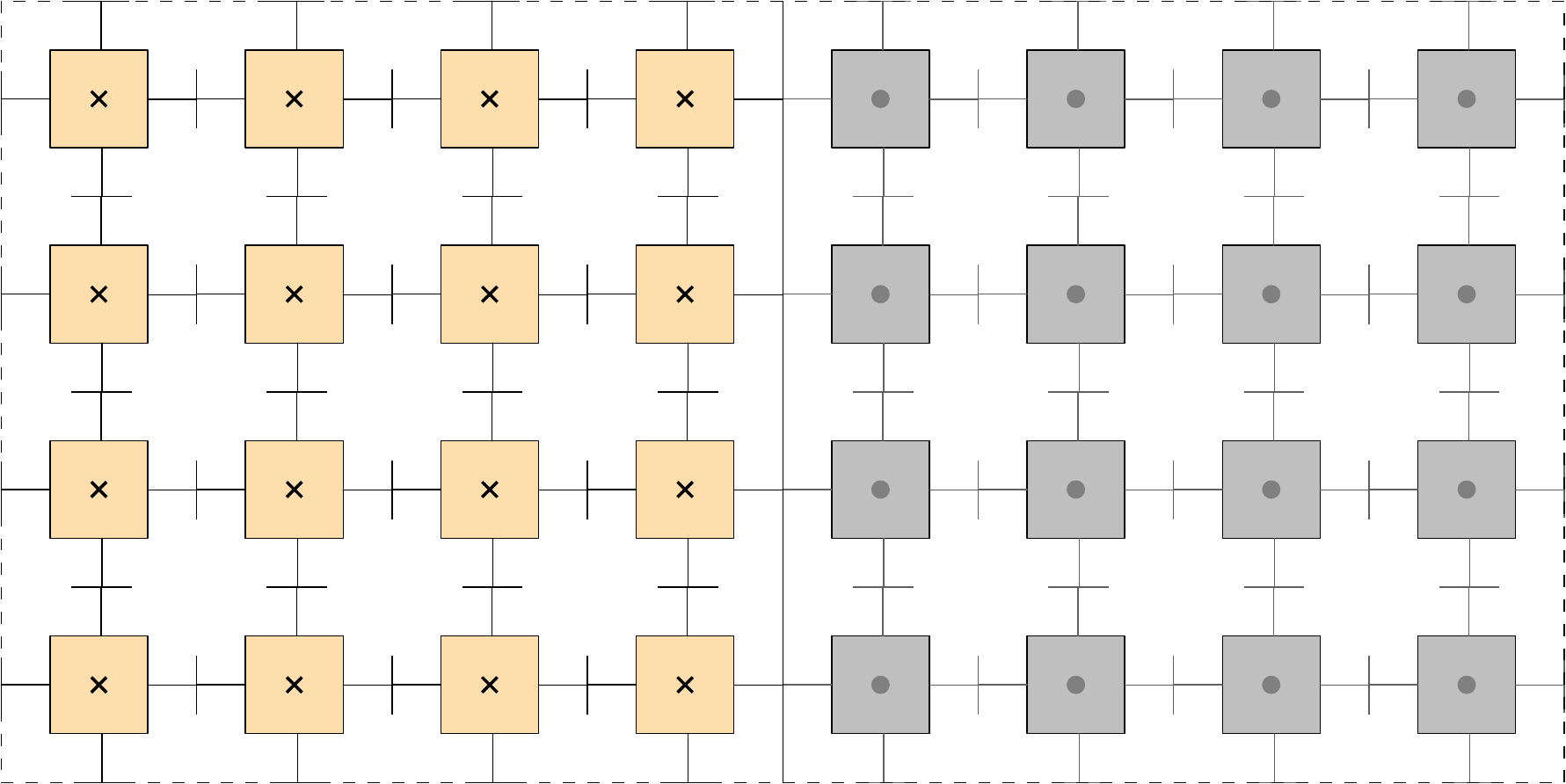} &
\includegraphics[height = 3.5cm]{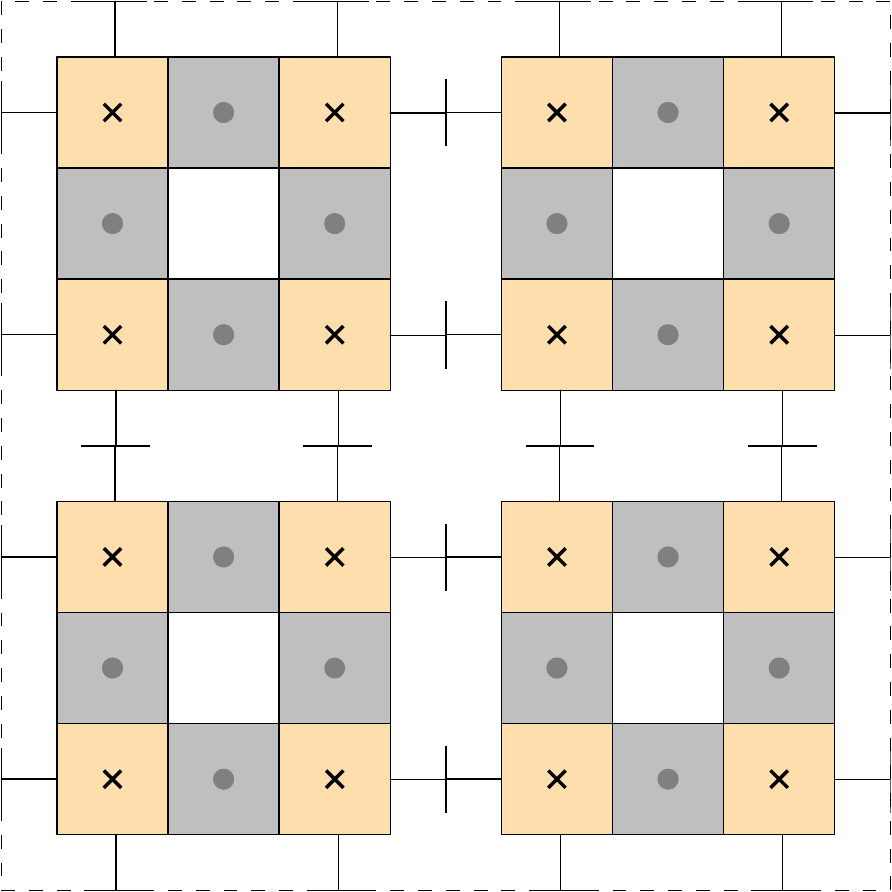}\\
\small (a) & \small (b)\\
\end{tabular}
\caption{\emph{Result of compressing two adjacent $4{\times}4$ modules into one lattice cell.}}
%\vspace{-3ex}
\label{fig_compression}
\end{figure}

%If a move requires $m>1$ this will be mentioned.
We first describe three basic module moves ({\em slide, compress, k-tunnel}) that are used in~\cite{cbots-cgta}.  We omit a detailed description of how to implement these moves in terms of individual atom operations.
A {\em compression} pushes one module $m_1$  into the space of an adjacent module $m_2$.
The atoms of $m_1$ literally fill the spaces between those of $m_2$ (see Fig~\ref{fig_compression}). Thus we see that two modules can occupy the same position in the module lattice. 
Any part of the robot attached to $m_1$ will be displaced by one unit along the same direction.   A {\em decompression} is the inverse of a compression.

\begin{figure}[htb]
\centering
%\begin{tabular}{c c}
\begin{tabular}{c @{\qquad\qquad} c}
\includegraphics[width = 5.0cm]{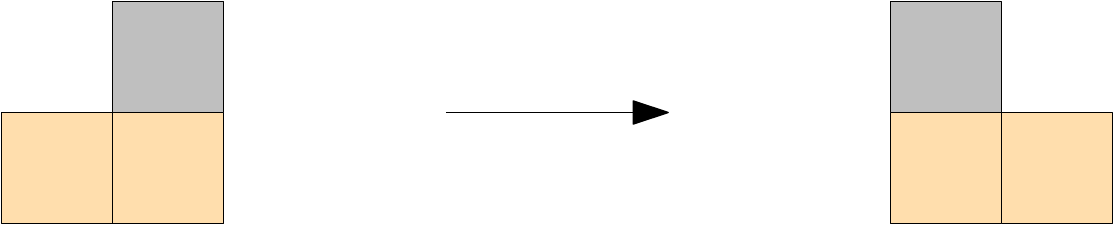} &
\includegraphics[width = 4.0cm]{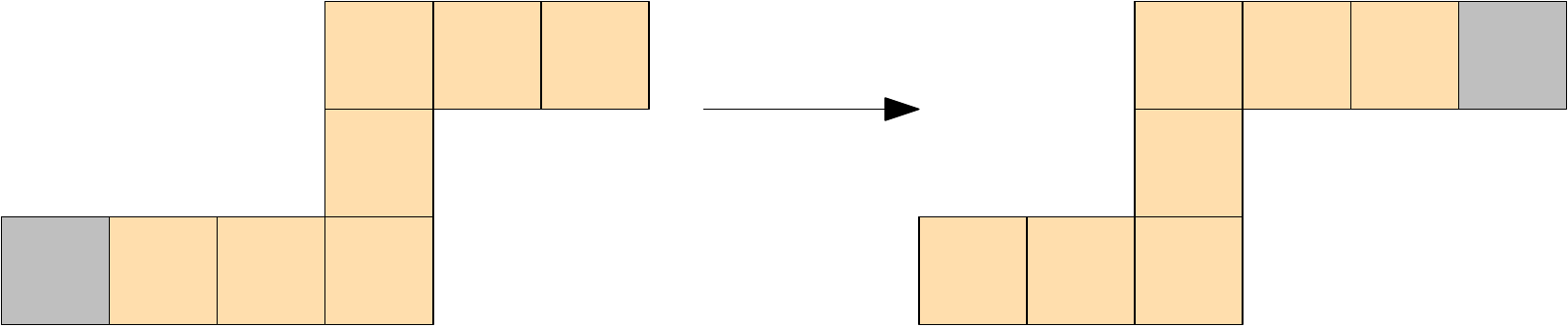}\\
(a) & (b)\\
\end{tabular}
\caption{\emph{(a) Slide move; (b) Tunnel move.}}
\label{figure_slide-and-tunnel}
\end{figure}

A \emph{slide} moves a module to an adjacent position,
using two substrate modules. See Fig.~\ref{figure_slide-and-tunnel}a.
%slides one step if it performs a slide move as shown in Fig.~\ref{slide_move}.
The \emph{$k$-tunnel move} compresses a leaf module
into the robot, and {\em decompresses} another module out into a leaf position. 
See Fig.~\ref{figure_slide-and-tunnel}b.
 An entire path of modules between the two leaves is involved in this move. Within each branch in this path, modules shift in the direction of the compression, and essentially transfer the compression to the next bend. Any modules attached to the branches will also shift. This issue is addressed later on.
%the highlighted squares are the two modules involved.
The parameter $k$ denotes the number of branches (or bends) in the path
between the two leaf modules.
The move takes $O(k)$ parallel steps, but in our uses $k$ will always be a small constant.

\noindent We now proceed to describe new basic moves that form the cornerstones of our reconfiguration algorithm.
%%%%%%%%%%%%%%%%%%%%%%%%%%%%%%%%%%%%%%
\subsection{Staircase Move}
%Let the origin of an axis-aligned rectangle be its lower-left corner.
The \emph{staircase move} transforms a rectangle of %dimensions
$k_1 \times k_2$ modules to one of dimensions $k_2 \times k_1$,
both sharing the same  lower-left corner~$C$.
Connectivity to the rest of the robot is maintained through the module at $C$,
and thus that module cannot move.
Without loss of generality, we can assume that $k_1\geq k_2$;
otherwise, we invert the sequence of operations described.

%%%\subsubsection{Move for $m \geq 2$}

%We have a rather easy staircase move which involves sliding all rows of the given rectangle simultaneously, to achieve a diagonal intermediate configuration (see Fig.~\ref{staircase}).  However this requires that the axis-parallel box of dimensions $k_1\times (k_1 + k_2 -2)$ at origin $C$ is empty of other  modules., and that $k_2\geq 3$.  Thus we proceed to describe a more complicated move that keeps the intermediate form within the bounding box of the initial and final rectangles.

%The staircase move applies slide moves to blocks of $2{\times}2$ \emph{basic modules}. The dimension of $k_2$ must be at least equivalent to two basic $2\times2$  modules, to permit the slide operation. Hence the requirement for $m\geq2$ if $k_2=1$.

\begin{figure}[htb]
\centering
\begin{tabular}{c @{\quad} c @{\quad} c @{\quad} c}
\includegraphics[height = 2.7cm]{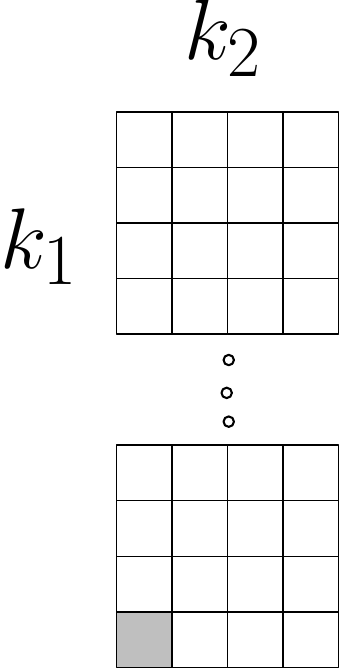} &
\includegraphics[height = 2.2cm]{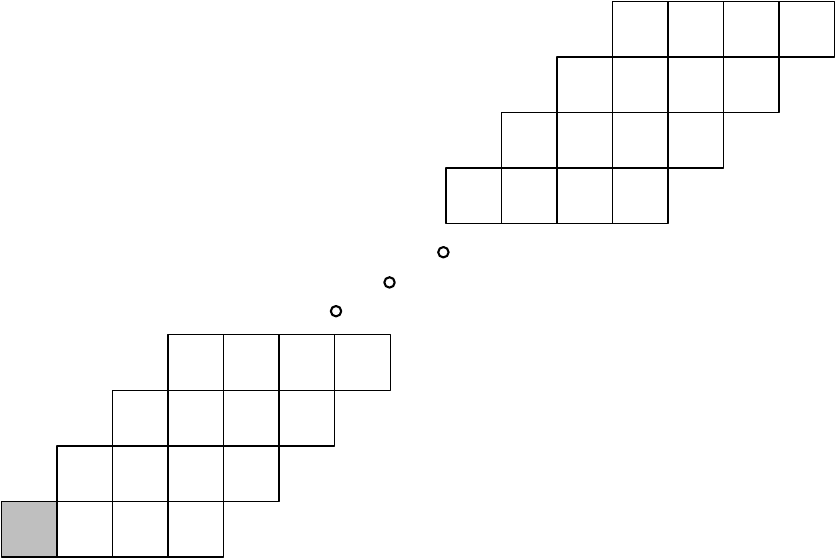} &
\includegraphics[height = 1.6cm]{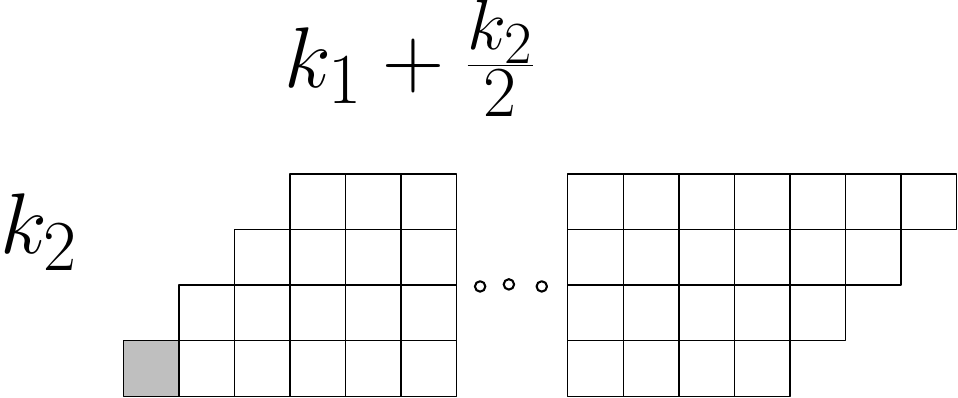} &
\includegraphics[height = 1.4cm]{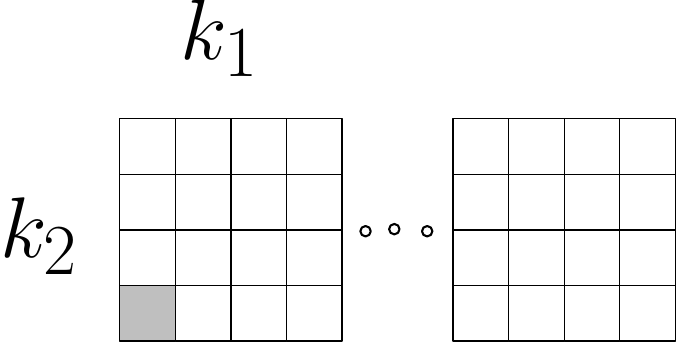}\\
(a) & (b) & (c) & (d)\\
\end{tabular}
\caption{\emph{Staircase move in three parallel steps. The shaded module maintains connectivity to the rest of the robot.}}
\label{staircase}
\end{figure}

First, we move every row of modules to the right using a slide move with respect to the row immediately below,
as in Fig.~\ref{staircase}(b).
Second, we move every column that does not touch the top or bottom border of
the bounding box down using a slide move, as in Fig.~\ref{staircase}(c).
Finally, we move every row to the left using a slide move,
as in Fig.~\ref{staircase}(d).
Note that the sliding motions of each step are executed in parallel. Also, each operation can be done at the atom-level, as was shown in Fig.~\ref{example}. Thus the move works even if $k_2 = 1$.
%Hence, a staircase move takes $O(1)$ parallel steps.

If we require that the transformation between rectangles takes place within
the bounding box of the source and target configurations,
we can modify the above procedure
%without much difficulty.
%This modification is omitted in the present version of this paper.
as follows.
%Assume for the moment that $k_2\geq 4$.
Ideally, every row of  modules would slide to the right  as illustrated in Fig.~\ref{staircase}(b). We partially perform  this step but do not slide the top rows that would emerge from the $k_1 \times k_1$ square originating at $C$.  Thus
we get a shape partially resembling Fig.~\ref{staircase}(b), but with a dimension $k_2$ square appended on top of it.
Now we slide the diagonal structure downwards, including the lower-right half of the extra square  (i.e., slide down everything below the diagonal starting at $C$).  At this point we have a shape similar to Fig.~\ref{staircase}(c) except that the empty region in the lower right corner of the figure is actually occupied, and there is a ``lower-left half'' of a dimension $k_2$ square sticking out above that corner.  Next, with a parallel slide to the left, we fill the empty region on the left but  create a symmetric empty region at the far right.  This leaves us with a rectangle exactly as we want it, except that at the far right we have a shape that is the lower left portion of a square (of height 2 times that of the rectangle).  We want to transport the top $1/4$ of that shapes' mass (i.e., the part that is above our target rectangle) into the the gap in the lower-right.  That's easily done with a few more slides. 

\subsection{Elevator Move}

\begin{figure}[htb]
\centering
\def\demiquad{\hskip0.8em}
\begin{tabular}{c @{\demiquad} c @{\demiquad} c @{\demiquad} c @{\demiquad} c}
\includegraphics[width = 2.2cm]{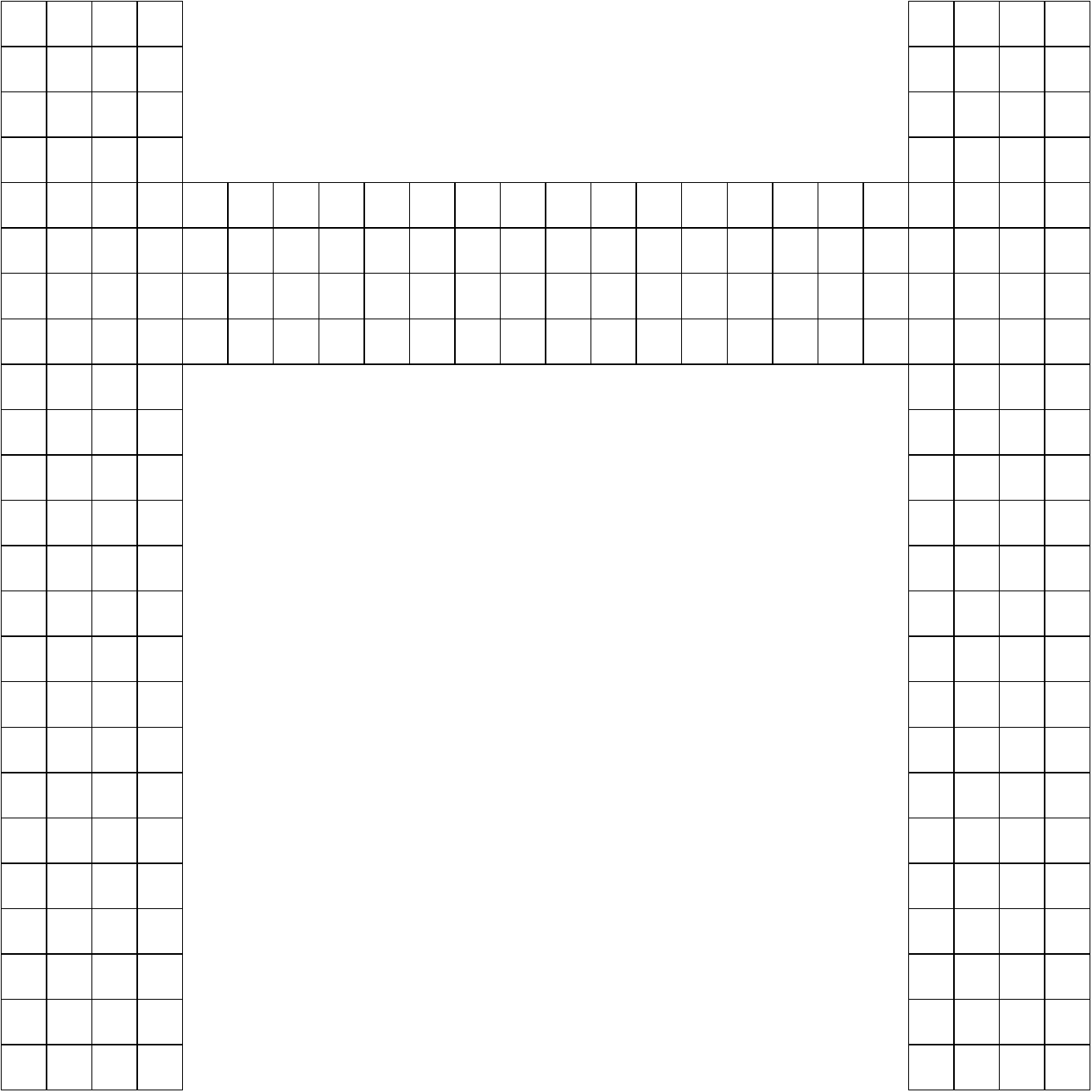} &
\includegraphics[width = 2.2cm]{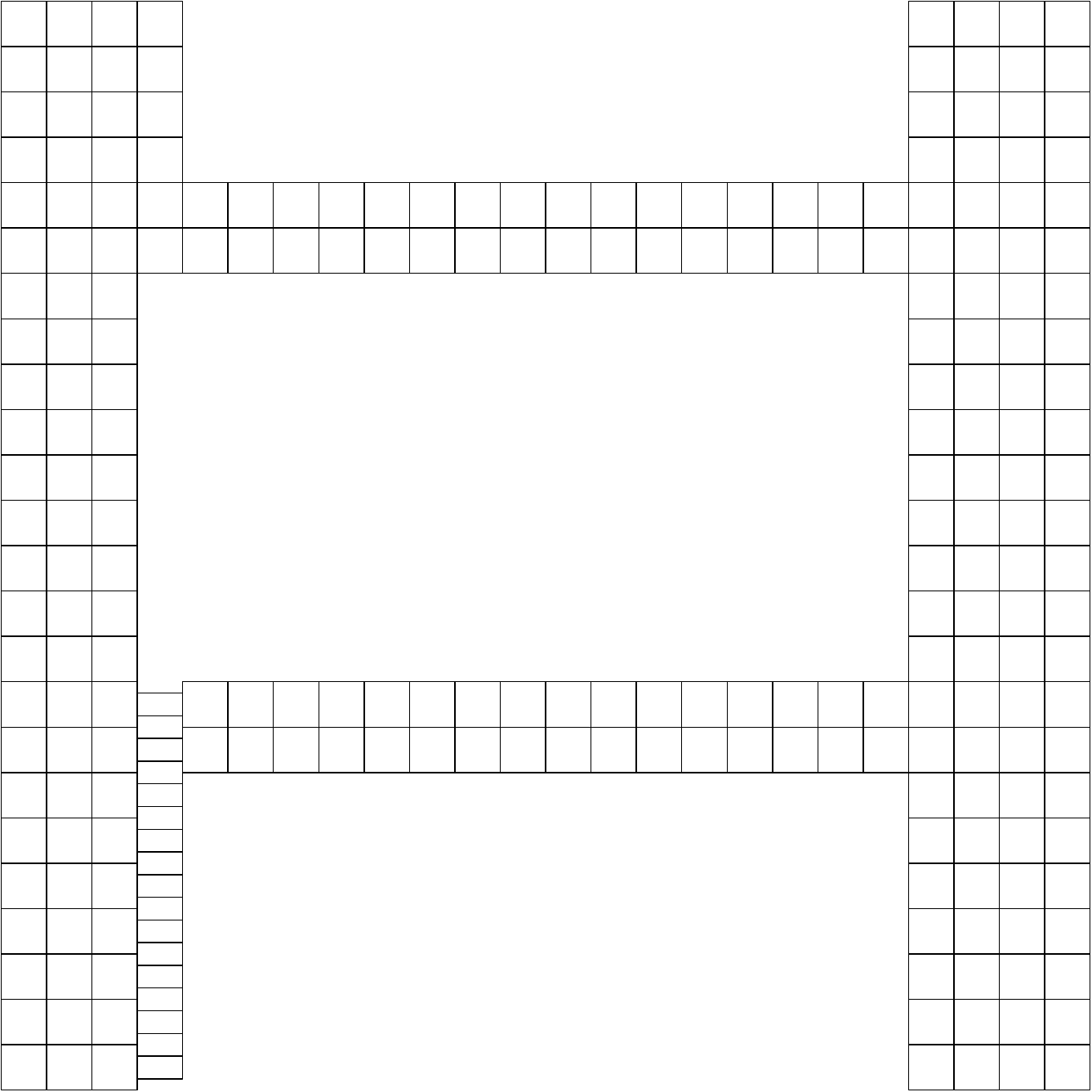} &
\includegraphics[width = 2.2cm]{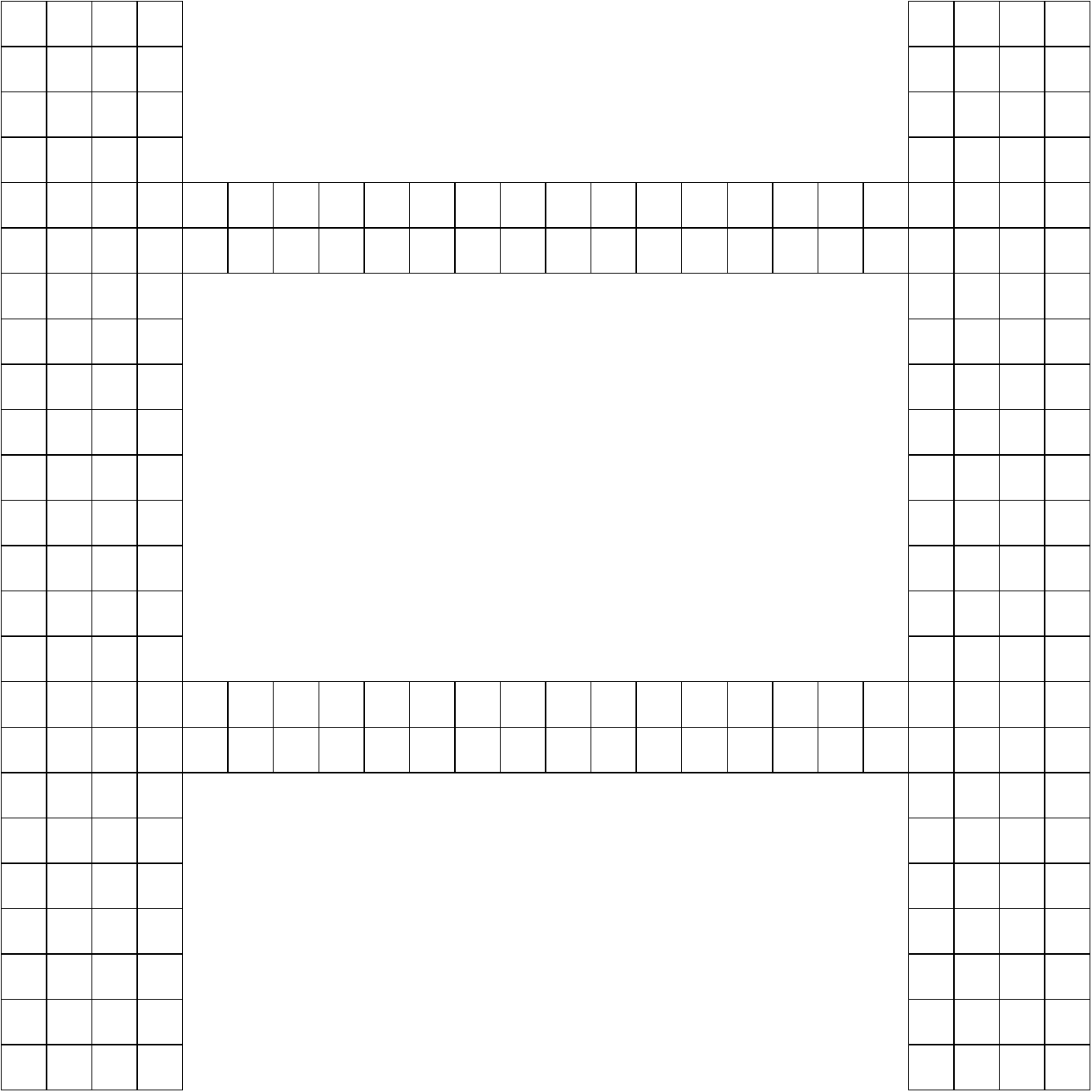} &
\includegraphics[width = 2.2cm]{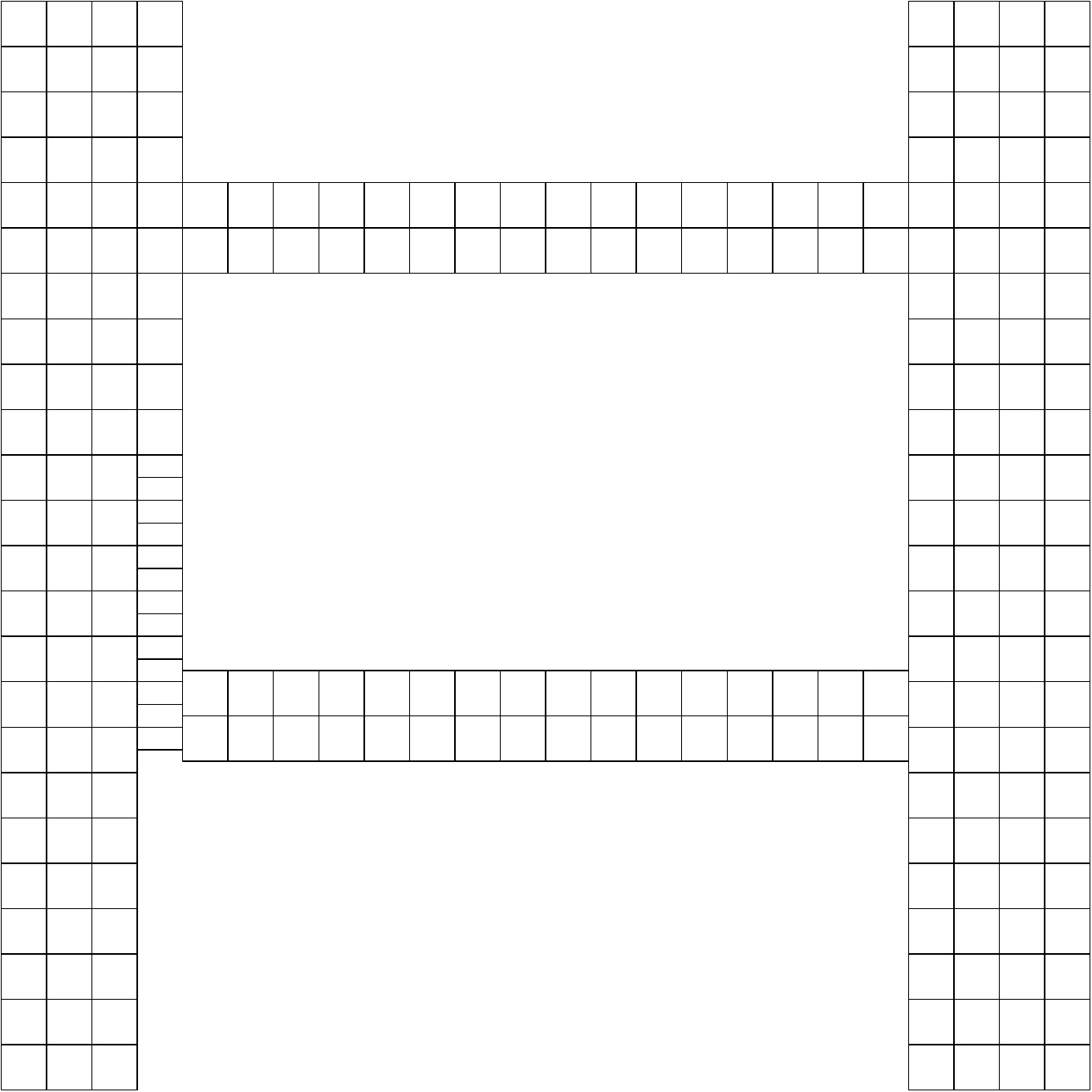} &
\includegraphics[width = 2.2cm]{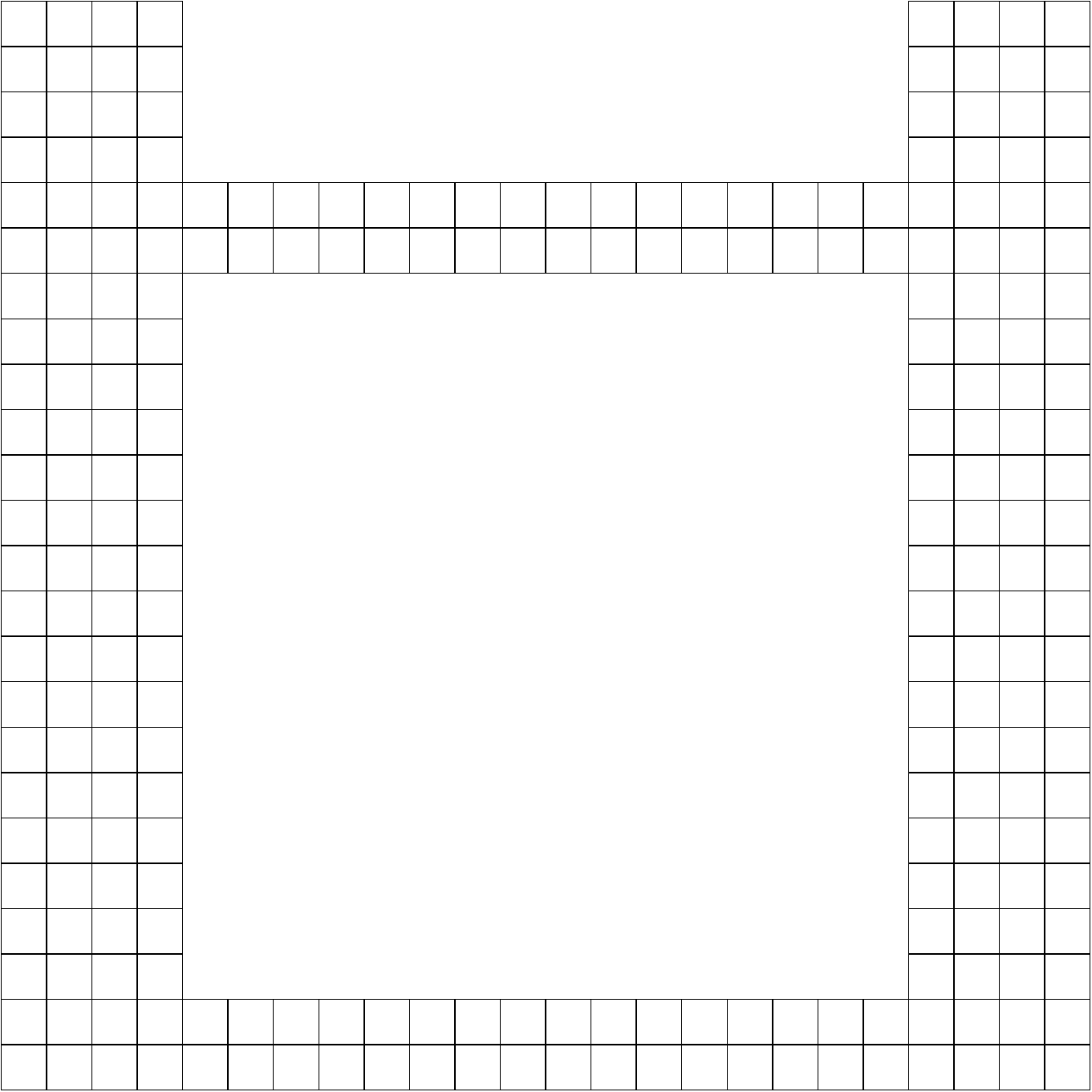}\\
(a) & (b) & (c) & (d) & (e)\\
\end{tabular}
\caption{\emph{Elevator move in $O(1)$ parallel steps.}}
\label{elevator}
\end{figure}

The \emph{elevator move} transports a rectangle of modules by $k$ units between two vertical strips, each at least two atoms wide.
Figure~\ref{elevator}(a) shows the initial configuration in which a rectangle
is to be transported vertically downward.
First we detach the top half $T$ of the rectangle from the bottom half $B$.
Furthermore, $B$ detaches from the vertical strip on the right.
Let $R$ be the rightmost vertical column of $k$ atoms along the left  strip,
together with the atoms to the left of~$B$.
We detach $R$ to its left, except at the very bottom, and detach $R$ above,
thus creating a corner with $B$.
Then we contract $R$ vertically, thereby pulling $B$ downward half way.
This is shown in Figure~\ref{elevator}(b), in which, however, we have let $R$ be a vertical column of modules instead of atoms, due to the large width of the shape.
Thus far, $T$ has maintained the connectivity of the robot.
Afterward, $B$ attaches to the right vertical strip and detaches from $R$,
which is now free to expand and re-attach to the top,
as in Figure~\ref{elevator}(c).
Now $R$ detaches from the bottom and contracts upwards.
It re-connects to $B$ at the bottom, as in Figure~\ref{elevator}(d).
In the last step, shown in Figure~\ref{elevator}(e),
$B$ detaches from the right side, and $R$ expands, thereby moving $B$
all the way to the bottom.
At this point, $B$  has reached its target position.
It now assumes the role of maintaining connectivity,
and the process can be repeated for $T$.  

\subsection{Corner Pop}
\begin{figure}[htb]
\centering
\def\demiquad{\hskip0.8em}
\begin{tabular}{c @{\demiquad} c @{\demiquad} c @{\demiquad} c @{\demiquad} c}
\includegraphics[width = 2.2cm]{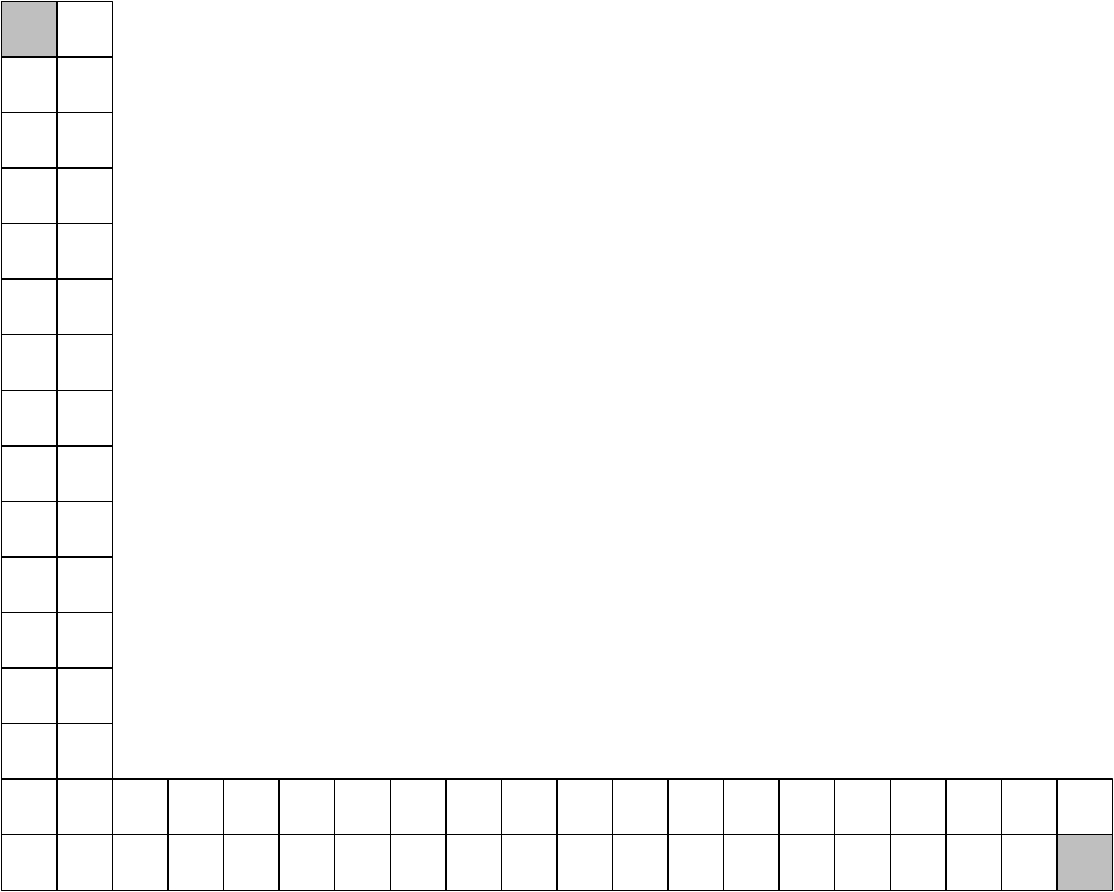} &
\includegraphics[width = 2.2cm]{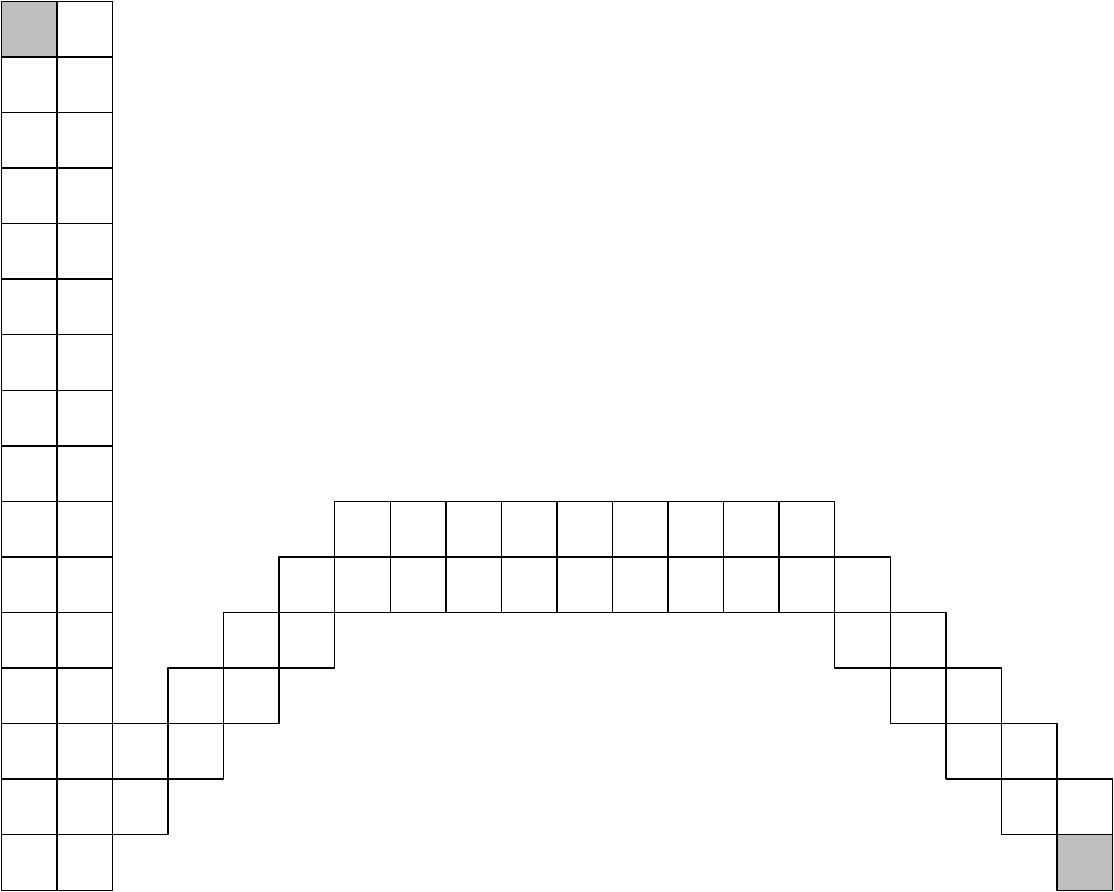} &
\includegraphics[width = 2.2cm]{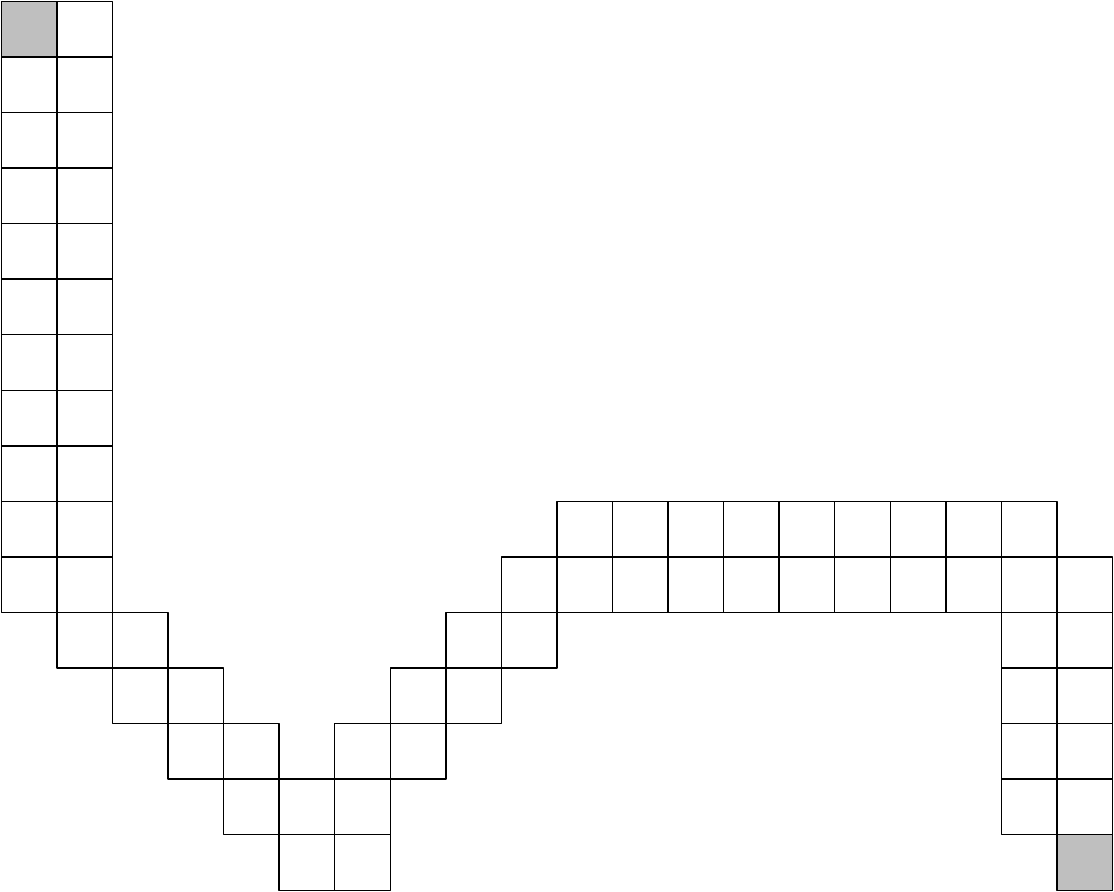} &
\includegraphics[width = 2.2cm]{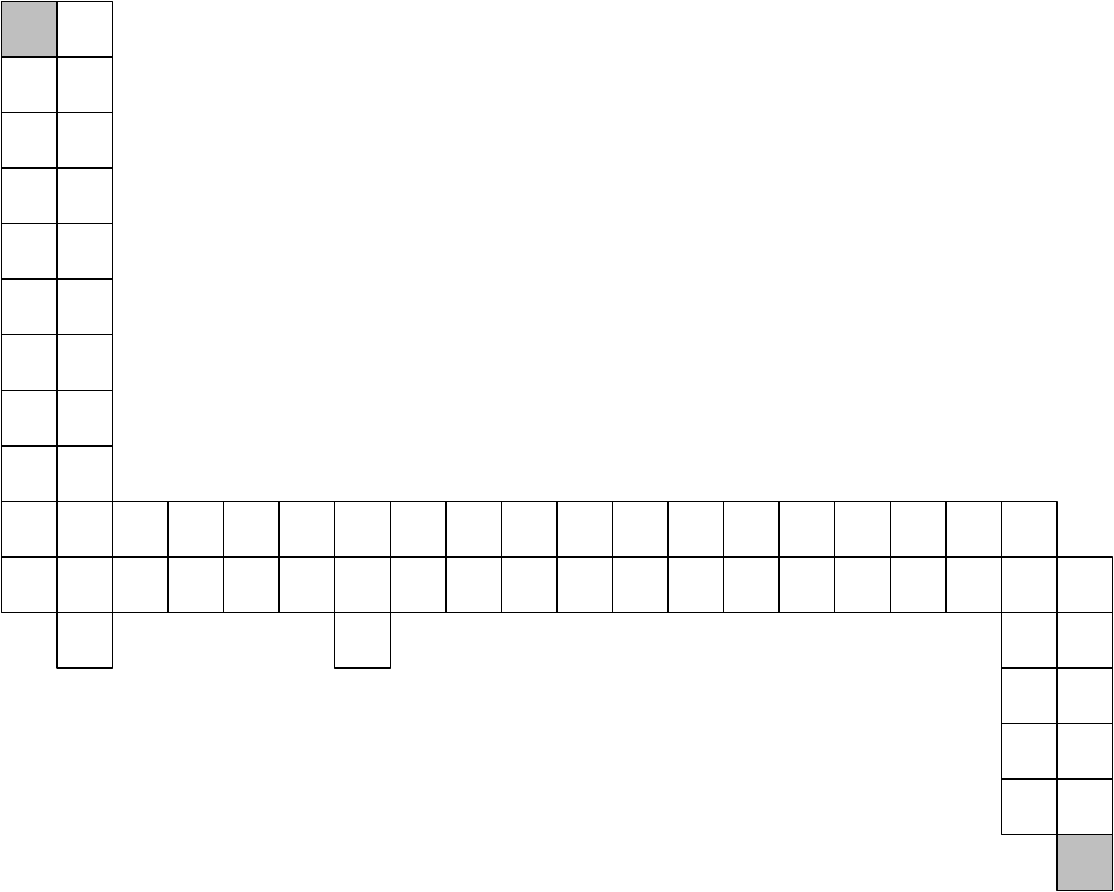} &
\includegraphics[width = 2.2cm]{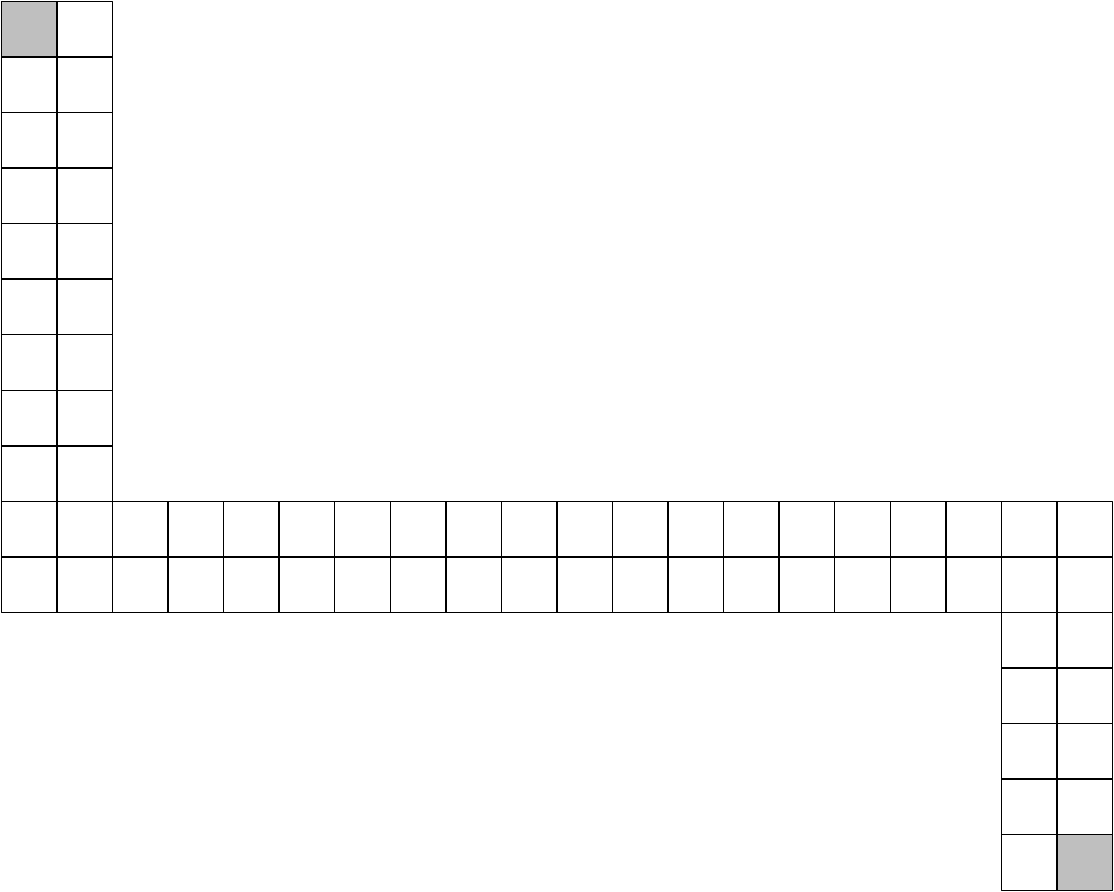}\\
(a) & (b) & (c) & (d) & (e)\\
\end{tabular}
\caption{\emph{Popping a corner in $O(1)$ parallel steps. The shaded modules maintain connectivity to the rest of the robot.}}
\label{corner}
\end{figure}

Consider a rectangle $R$ of $k_1 \times k_2$ module units,
 where without loss of generality $k_1 \leq k_2$.
Let $R$ be empty except for a strip $V$ of modules on its left border
and a strip $H$ along the bottom.
The strips form a corner, as shown in Figure~\ref{corner}(a).

The \emph{corner pop} moves the modules in $R$ to the upper and right borders
of~$R$.
%; thus the corner ``pops'' to the top-right.
During this corner pop, the modules at the top-left and bottom-right corners
of $R$ do not move.
It is assumed that only these positions connect to modules outside~$R$.
Thus, this operation preserves the connectivity of the robot.

%\subsubsection{Move for $m \geq 2$}

Assume that we have $4{\times}4$ modules, so that we can use our easy staircase operation.
We first create two staircases of height $k_1/2$ at the two ends
of $H$, as in Figure~\ref{corner}(b). This shifts the middle of $H$ upward.
Next, we use the lower half of $V$ to create a staircase of width $k_1/2$. Simultaneously, the rightmost staircase of $H$
also moves so that it ends up on the right border of $B$,
as in Figure~\ref{corner}(c).
%This creates two staircases of size $k_1/2$, each in the middle of the horizontal strip.
We move the two remaining staircases upward, as in Figure~\ref{corner}(d).
Finally, we clean up the two extra atoms that protrude using a $2$-tunnel move and a $3$-tunnel.

\subsection{Parallel Tunnel Move}

The \emph{parallel tunnel move} takes as input a horizontal row $H$ of
modules together with, on the row immediately above,
several smaller horizontal components that have no other connections.
The top components are absorbed into $H$, after which $H$ extends horizontally.
Alternatively, the absorbed mass can be pushed out anywhere else on top
of~$H$, provided the target space is free.
This move allows us to merge an arbitrary number of strips in the top row
in $O(1)$ time.

The idea is to take all odd lattice positions along $H$ and perform $1$-tunnel moves,
i.e., absorb modules from above and compress them under even positions.
Then decompressing them all in parallel just expands $H$ horizontally.
Any modules remaining on top will shift over during the expansion, since they are attached to $H$.
A gap will remain to the right of each such module, so we can repeat  one more time to complete the move.

\begin{figure}[htb]
\centering
\def\demiquad{\hskip0.6em}
\begin{tabular}{c 
@{\demiquad} @{\demiquad} c}  
\includegraphics[height = 3.5cm]{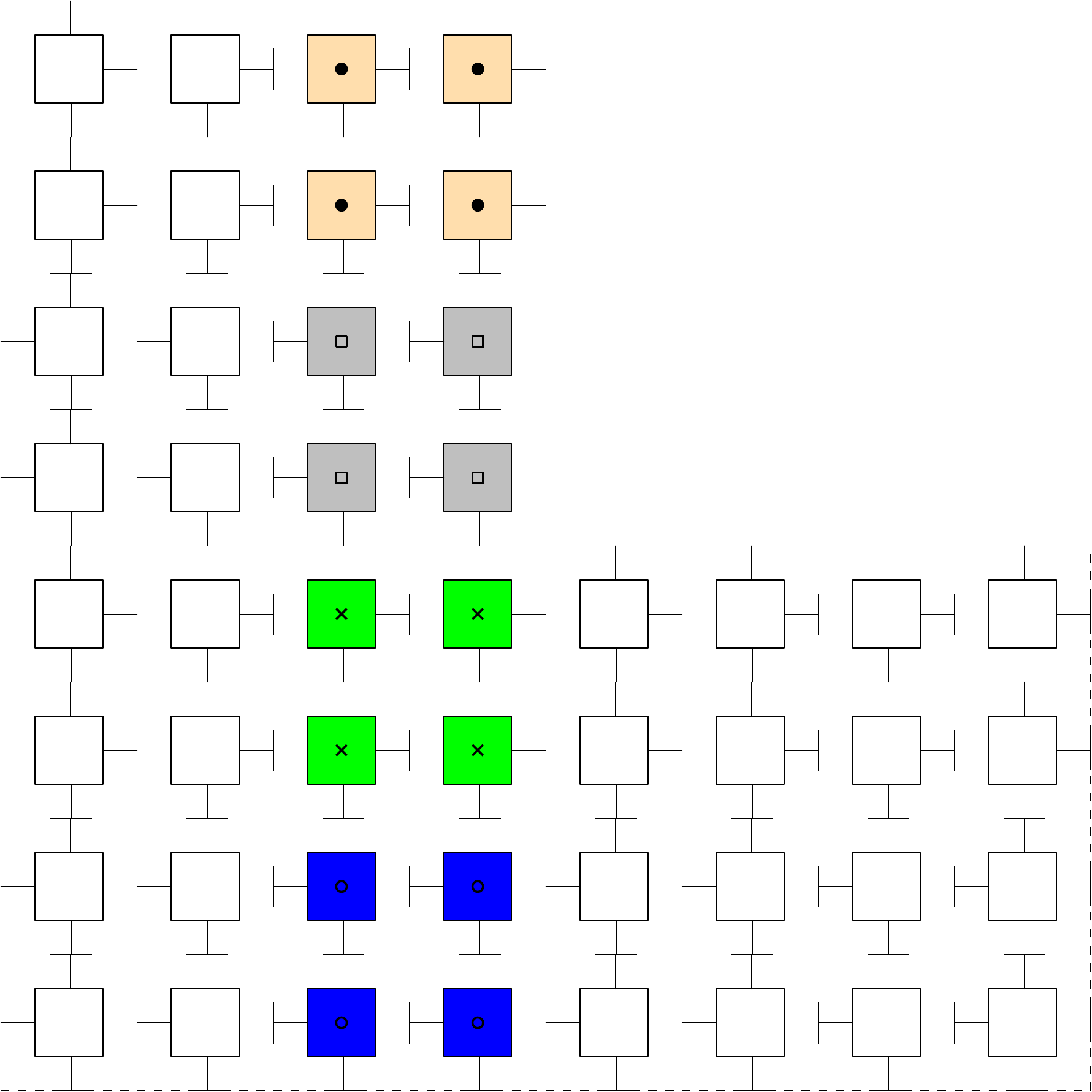} &
\includegraphics[height = 3.5cm]{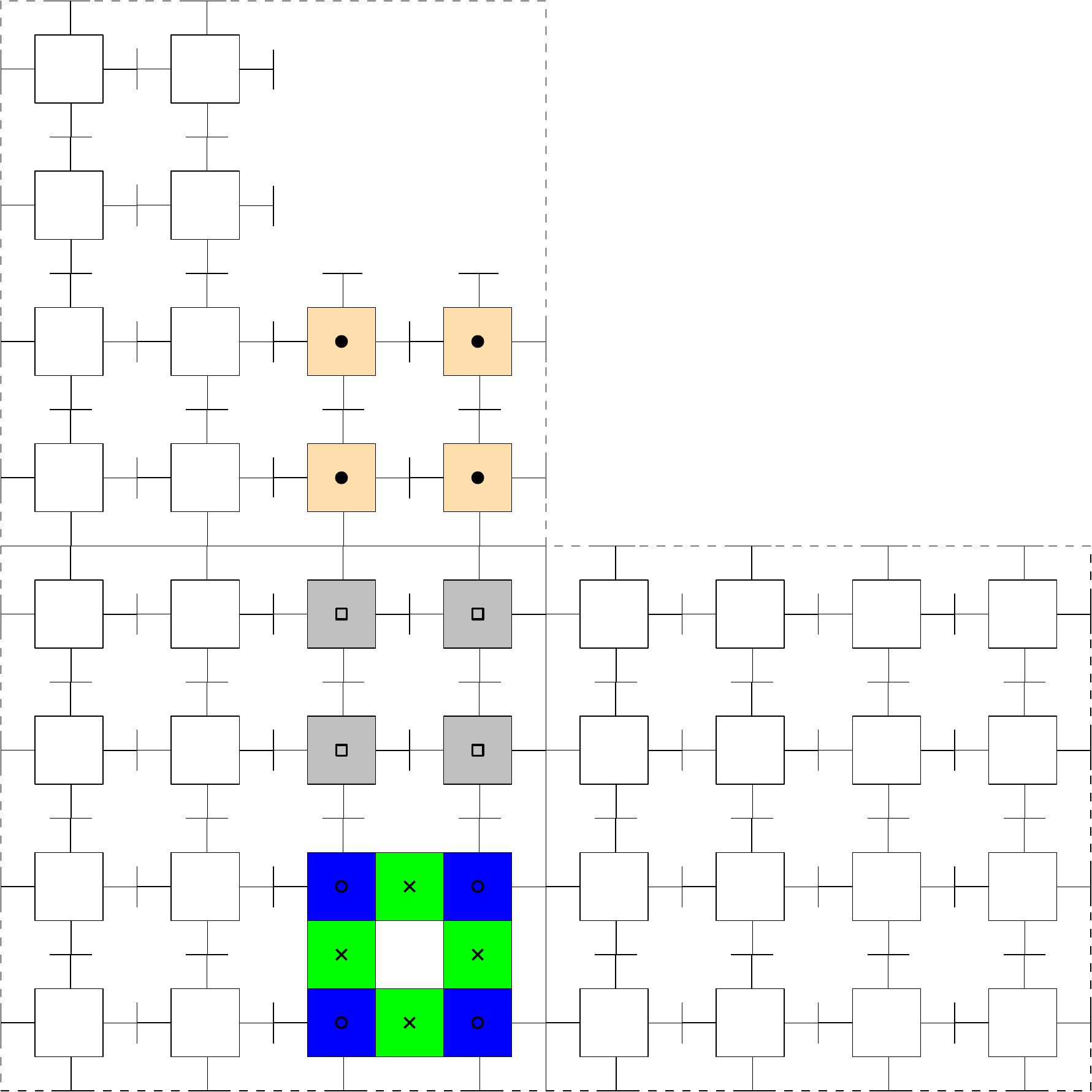} \\
(a) & (b)\\
\includegraphics[height = 3.5cm]{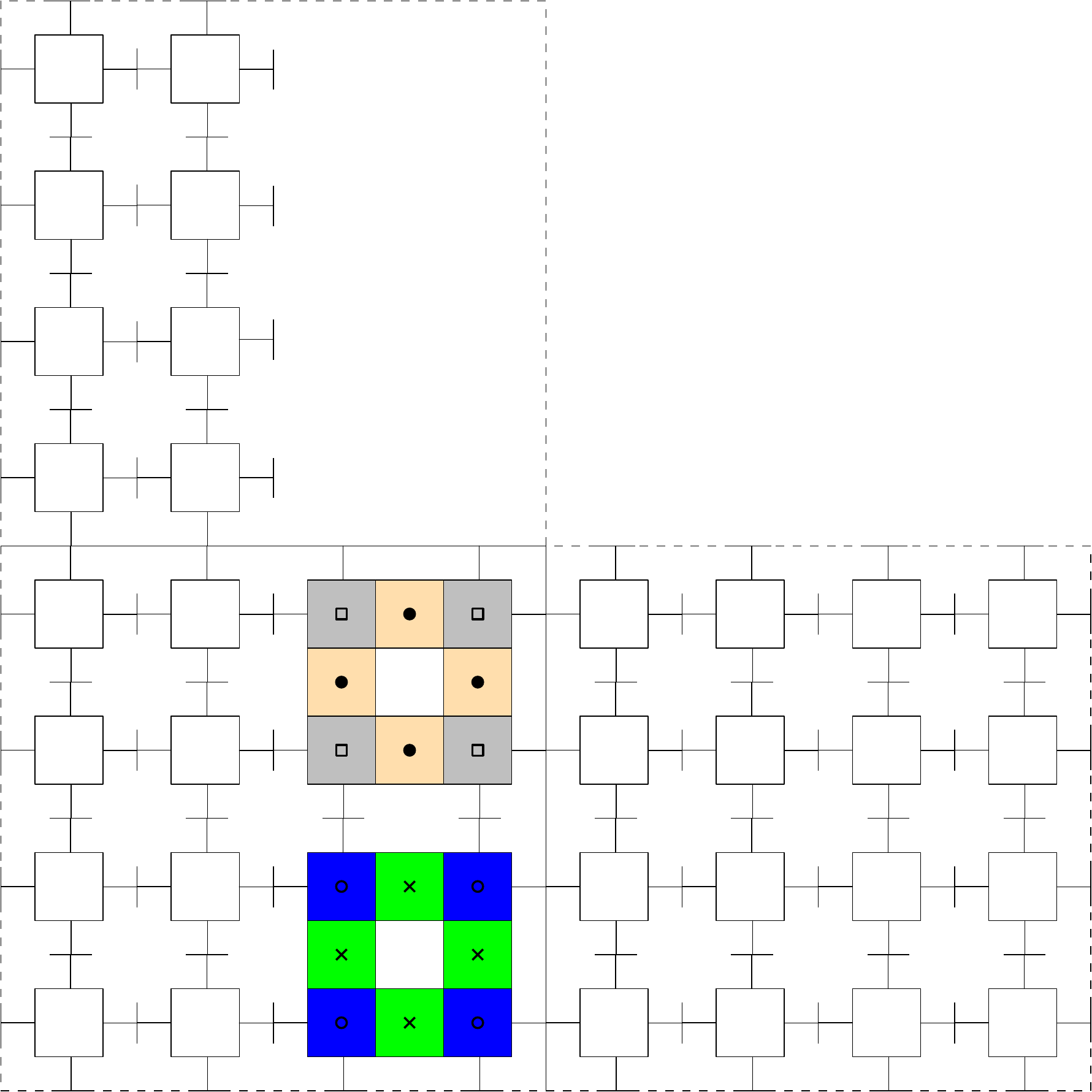} &
{\hskip 0.92cm}\includegraphics[height = 3.5cm]{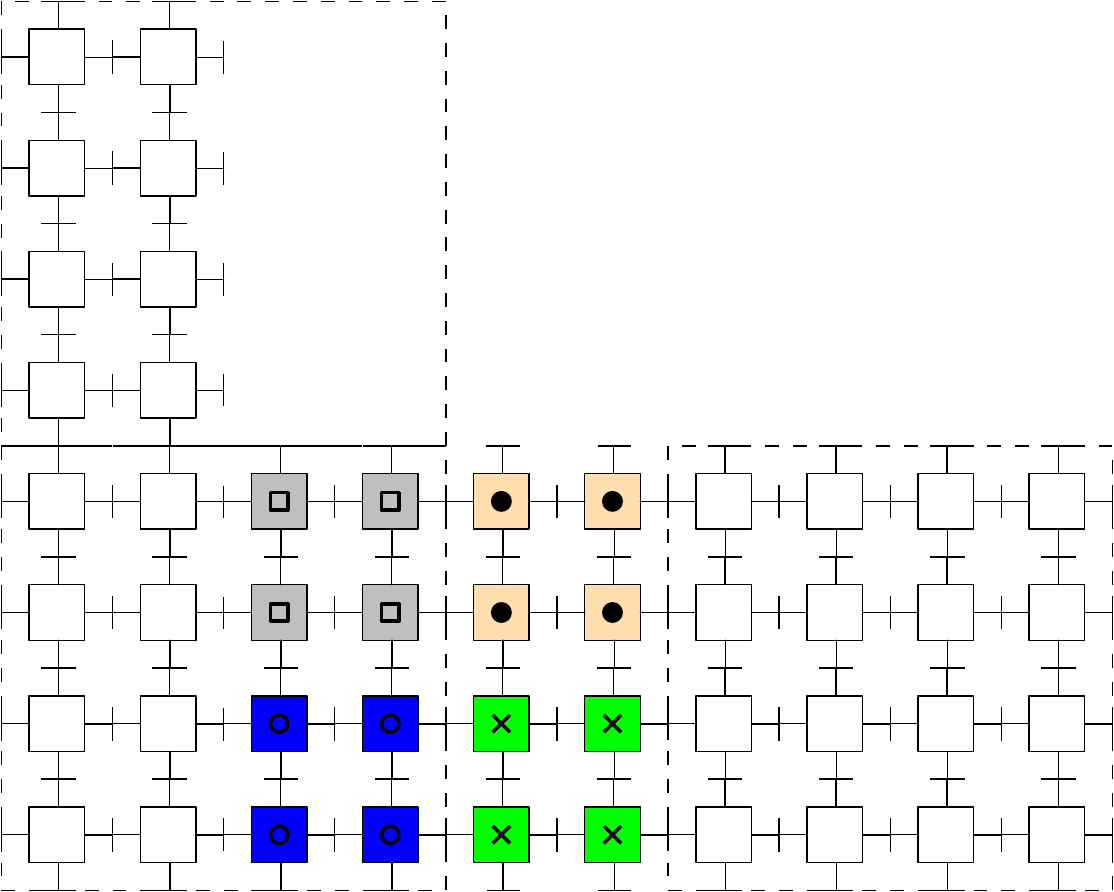}\\
(c) & (d)\\
\end{tabular}
\caption{\emph{Parallel tunnel move. Three $4\times4$ modules are involved.}}
\label{parallel_tunnel}
\end{figure}

Figure~\ref{parallel_tunnel} illustrates half of the absorption of one $4{\times}4$ module into $H$.
Note that groups of 4 atoms move separately (they can be considered to be temporary smaller modules).

\begin{comment}
\begin{figure}[htb]
\centering
\def\demiquad{\hskip0.6em}
\begin{tabular}{c @{\demiquad} @{\demiquad} c@{\demiquad} @{\demiquad} 
 c @{\demiquad} @{\demiquad} c}
\includegraphics[height = 2.5cm]{Figs/parallel_tunnel_detail_1.pdf} &
\includegraphics[height = 2.5cm]{Figs/parallel_tunnel_detail_2.pdf} &
\includegraphics[height = 2.5cm]{Figs/parallel_tunnel_detail_3.pdf} &
\includegraphics[height = 2.5cm]{Figs/parallel_tunnel_detail_4.pdf}\\
(a) & (b) & (c) & (d)\\
\end{tabular}
\caption{\emph{Parallel tunnel move. Three $4\times4$ modules are involved.}}
\label{parallel_tunnel}
\end{figure}
\end{comment}
%%%%%%%%%%%%%%%%%%%%%%%%%%%%%%%%%%%%%%%%%%%%%%
%%%%%%%%%%%%%%%%%%%%%%%%%%%%%%%%%%%%%%%%%%%%%%
%%%%%%%%%%%%%%%%%%%%%%%%%%%%%%%%%%%%%%%%%%%%%%
%%%%%%%%%%%%%%%%%%%%%%%%%%%%%%%%%%%%%%%%%%%%%%
%%%%%%%%%%%%%%%%%%%%%%%%%%%%%%%%%%%%%%%%%%%%%%
%%%%%%%%%%%%%%%%%%%%%%%%%%%%%%%%%%%%%%%%%%%%%%

\section{Reconfiguration}
\label{k=32}
In this section we show how to reconfigure a given robot to a canonical form with $O(\log n)$ parallel steps.
Here we assume that the initial and final configurations of the robot consist of blocks of $32{\times}32$  atoms. However we will split blocks to use  modules of $4{\times}4$ atoms in the intermediate configurations.
%Let a \emph{thin ring} (respectively \emph{thin blob, thin  sparse cell}) have a thickness of 4 atoms (2 basic  modules).
Recall that
%We consider the definition of  boundary to be determined by modules, not the larger initial blocks.
%Hence
the boundary has a width of four atoms.
%correspond to the shape of a thin ring, not a ring of $8\times8$  modules.

Our divide-and-conquer algorithm proceeds as follows.
Let the initial robot be placed on a grid of unit blocks (of $32{\times32}$ atoms).
On this grid we construct a minimal square cell of side length $2^h$ that contains the initial robot (length is measured in block units).
We recursively divide the cell into four subcells of length $2^{h-1}$.
As a base case, we take subcells of $2{\times}2$ blocks (i.e., containing $16{\times}16$ module lattice positions).

In parallel, we reconfigure each subcell within the same recursive depth, so that the resulting shape is easy to handle.
Thus, by merging subcells, in $O(\log n)$ iterations we will have created a simple shape in our original square.
Consider a cell $M$.  We will use the inductive hypothesis that after merging its subcells, $M$ will become a ring if there are enough modules, or sparse otherwise.  Furthermore,
if two points on the boundary of $M$ were initially connected, the new configuration will ensure connectivity via the shortest path through its boundary.

In the base case of our induction, $M$ has length 2. Thus we have to merge four subcells, each of which is empty or full.
We will obtain a ring if there is at least one full subcell.  One such subcell contains $64$ modules, which suffice to cover the boundary of $M$.  Reconfiguration can be done by tunneling each interior module iteratively (or by the lemmas that will follow).  Thus our hypothesis is preserved.

\begin{figure}[htb]
\centering
%\begin{tabular}{c c}
\begin{tabular}{c @{\qquad} c}
\includegraphics[height = 3.0cm]{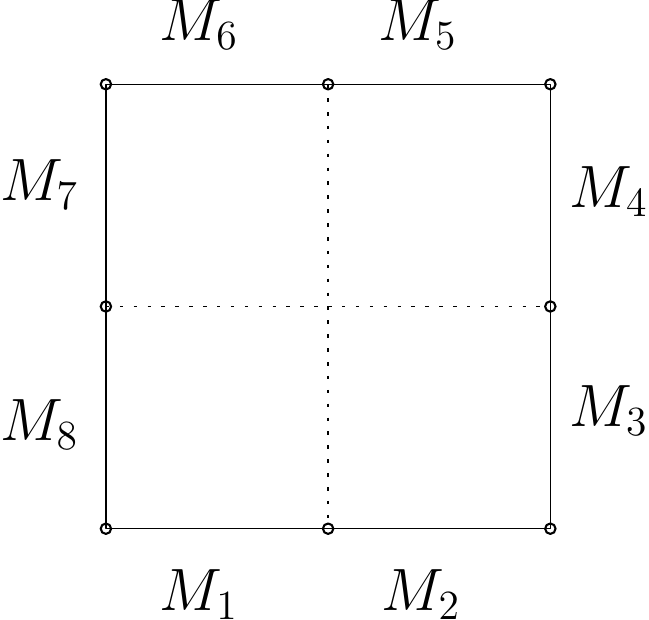} &
\includegraphics[height = 2.85cm]{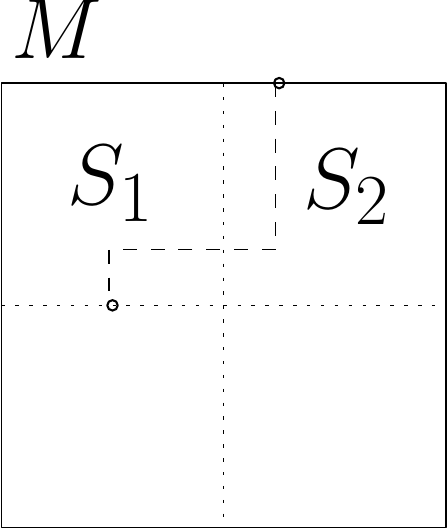}\\
(a) & (b)\\
\end{tabular}
\caption{\emph{Connectivity issues, in Lemmas~\ref{enough4ring} and~\ref{2sparse}.}}
\label{figure_connectivity}
\end{figure}

%\begin{lem}
%\label{2forms}
%The configuration of any cell $M$ can be substituted with a  ring or  sparse configuration, without disrupting connectivity between modules that possibly exist outside $M$.
%\end{lem}
%\begin{proof}
\begin{lem}
\label{enough4ring}
Consider a cell $M$.
If any subcell of $M$ contained a backbone
%of $32\times32$-atom blocks
 in the original configuration, then there are enough  modules to create a ring in $M$.
There are also enough modules if a path originally connected two subcell sides that belong to the boundary of $M$ but are not adjacent.
\end{lem}
\begin{proof}
Consider the eight exterior sides of subcells of $M$ as shown in Figure~\ref{figure_connectivity}(a). Let each of the sides $M_i$ have length $c$ (i.e., $c$ modules fill the side of a subcell). The total number of modules in the boundary of $M$ is $8c-4$.  A subcell backbone contains at least
$8c$ modules and therefore suffices to cover the boundary.

Without loss of generality, suppose that a path begins on $M_1$ and ends at any side other than $M_1,M_8,M_2$. Then we have enough modules to make a ring in $M$, by similar counting as above.  In fact to avoid having enough modules, such a path would have to remain within the lower two subcells.
\end{proof}

%\begin{figure}[htb]
%\centering
%\includegraphics[height = 2.0cm]{Figs/fig_connectivity.pdf}
%\label{figure_connectivity}
%\end{figure}

\begin{lem}
\label{2sparse}
Let $S_1$ and $S_2$ be adjacent sparse subcells at the top of cell $M$.  In the original robot, there can be no path from the top border of $M$ to the other subcells
%(see Figure~\ref{fig-2sparse}).
(see Figure~\ref{figure_connectivity}(b)).
\end{lem}
\begin{proof}
A path from the top to the middle of $M$ in the initial robot would contain enough modules to make both $S_1$ and $S_2$ rings.  By the pigeon-hole principle, one of the two subcells cannot be sparse.
\end{proof}
%\begin{figure}[htb]
%\centering
%\includegraphics[height = 2.0cm]{Figs/Illustration_Lemma6.pdf}
%\caption{The dashed path will cause $M_1$ or $M_2$ to become a ring.}
%\label{fig-2sparse}
%\end{figure}

\begin{comment}
So assume that there is only connectivity as described above. In other words, $M$ contains  disjoint subgraphs, each of which does not contain a backbone.  Also each subgraph passes through at most two subcells.
Specifically, a subgraph is defined as follows:  let $A$ be a connected component of robot in $M$, touching at most two consecutive external subcell sides $M_i,M_{i+1}$. Let $a_1,a_2$ be the extremal points of contact of $A$ on  $M_i,M_{i+1}$.  Other connected components may be ``covered'' by $A$, which implies that their extremal points are nested between $a_1,a_2$. Assume that $A$ itself is not covered in this way, i.e., it is the maximum of such nested components, all of which form one of our disjoint subgraphs.

Let $M_A$ be the path from $a_1$ to $a_2$ along $M_i,M_{i+1}$. Let $E_A$ be the path from $a_1$ to $a_2$ in $A$, which together with $M_A$  encloses the rest of $A$ (and any other connected component covered by $A$).  The modules in $E_A$ suffice to cover $M_A$ because the latter is a shortest path between $a_1$ and $a_2$.  In fact they cover all connection points of the associated subgraph on the boundary of $M$.
\end{comment}

%%%%%%%%%%%%%%%%%%%%%%%%%%%%%%%%%%%%%%%%%%%%%

\begin{lem}
\label{mergeblob}
All side-branches along the common border of two cells that are rings or sparse can be merged into at most two pieces per side, with $O(1)$ moves. Furthermore each side-branch touches one end of the border.
\end{lem}
\begin{proof}
If one cell is a ring then the other side can use it as a platform for a parallel tunnel move that will merge its side-branches into one piece.
Otherwise, for each connected component of side-branches (of which there are at most two; one per corner) do the following.

Denote the two sides of the border by $A$ and $B$. Absorb as much as possible from $A$ to $B$ by sliding modules from $A$ across the border into vacant module lattice positions.  Thus the component has one side-branch in $B$. Shift (parallel tunnel) the remainder of $A$ towards the corner that the connected component attaches to, using $B$ as a platform.  Thus  $A$ becomes one side-branch. Now (either by a pop or by parallel-tunnel) bring back material from $B$ to $A$ to restore the original numbers in each cell. Thus each connected component consists of at most one side-branch from $A$ and one from $B$.
\end{proof}

\begin{figure}[htb]
\centering
\begin{tabular}{c @{\qquad} c @{\qquad} c}
\includegraphics[height = 1.7cm]{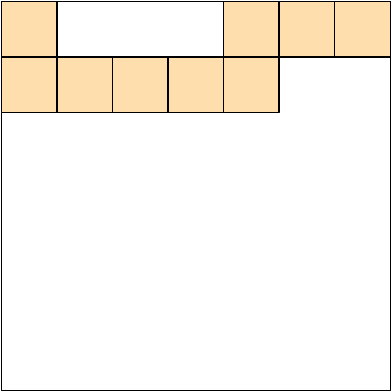} &
\includegraphics[height = 1.7cm]{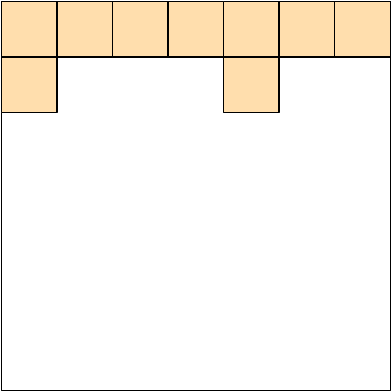} &
\includegraphics[height = 1.7cm]{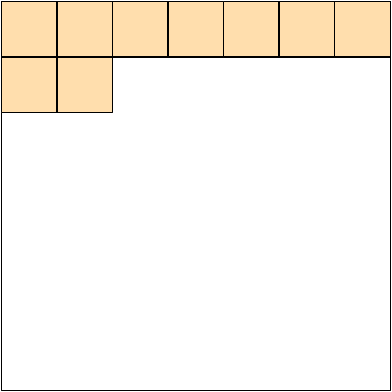}\\
(a) & (b) & (c)\\
\end{tabular}
\caption{\emph{Absorbing a near-boundary branch into the boundary of a cell.}}
\label{fig-lem-into-boundary}
\end{figure}

\begin{lem}
\label{into-boundary}
Suppose $B$ is a boundary side of a cell that has been processed according to Lemma~\ref{mergeblob}.
Let  $A$ be a branch that is  in the near-boundary adjacent to $B$, and has no connectivity purpose. We can absorb $A$  into $B$, or $B$ can be filled, with $O(1)$ moves.
\end{lem}
\begin{proof}
By Lemma~\ref{mergeblob}, $B$ contains at most two side-branches, each attached to a corner.
If no modules in $B$ are adjacent to $A$, we can use a $1$-tunnel to move one node (endpoint) of $A$ into the position in $B$ that is adjacent to the other node of $A$.   Then the rest of $A$ can slide into $B$.
Otherwise, if $A$ is adjacent to a side-branch in $B$, as in Figure~\ref{fig-lem-into-boundary}(a), we do the following.
Absorb parts of $A$ into empty positions of $B$, as in Figure~\ref{fig-lem-into-boundary}(b).  Thus we create a side-branch $B_1$ which can be used as a platform to be extended by performing a parallel tunnel move on what remains of $A$.  If the extension causes $B_1$ to reach a corner or join to another side-branch in $B$, then $B$ is full; see Figure~\ref{fig-lem-into-boundary}(c).
\end{proof}
For sparse cells, by repeatedly applying Lemma~\ref{into-boundary} and staircaising the remainder of $A$ to the near-boundary side adjacent to $B$, we obtain the following:
\begin{cor}
\label{rid-blob}
If a branch $A$ is positioned in the near-boundary of a sparse cell, either $A$ can be fully absorbed into the boundary, or the cell will become a ring.
\end{cor}

Let a \emph{merged cell}  contain four subcells that satisfy our induction hypothesis.  That is, they are either rings or sparse, and connectivity is ensured via shortest paths along their boundaries.
A merged cell becomes \emph{well-merged} if it is reconfigured to satisfy the induction hypothesis.

\begin{lem}
\label{3-4rings}
Let $M$ be a merged cell containing three or four subcell rings.  Then $M$ can become a ring using $O(1)$ moves. Thus $M$ becomes well-merged.
\end{lem}
\begin{proof}
%Omitted due to space restrictions.

%\noindent {\em Sketch:} The outer structure of the desired ring is either in place or can be completed easily.  Following this, all that remains is to organize/merge  the interior modules of the subcells.
First consider the case where $M$ consists of four rings. The outer ring structure is already in place.
Use elevator moves to transfer any internal material from the top two subcells as far down as possible.
The resulting interior structure on either side of the internal vertical subcell borders is a monotone terrain with at most two height changes. Furthermore each height change is by one unit.
The internal borders can be staircased and incorporated into these terrains, which can then be merged easily with parallel tunneling and sliding.
There still may be a large height difference between the left and right sides. In other words, one side has an extra rectangle of modules. To even things out,  we slide half of the rectangle over and then use two staircase moves.

Second consider the case where $M$ consists of three rings and a sparse subcell $S$. We can quickly modify the ``direction'' of the interior structure in any ring with a couple of staircase moves, so
 without loss of generality let $S$ be at the top-left  of $M$.
 If the interior sides of $S$ contain side-branches, we can staircase them to the near-boundary and then apply Corollary~\ref{rid-blob}, to redistribute them to exterior sides or obtain a fourth ring.
 In the former case, we are free to staircase the full sides of subcells adjacent to $S$ in order to fill the boundary of $M$ (again using Corollary~\ref{rid-blob}).  Further redistribution is nearly identical to the four ring case.
\end{proof}

%%%%%%%%%%%%%%%%%%%%%%%%%%%%%%%%%%%%%%%%%%%%%%%

\begin{lem}
\label{2rings}
If exactly two subcells of a merged cell $M$ are  rings, then $M$ can become well-merged using $O(1)$ moves.
\end{lem}
\begin{proof}
If the two sparse subcells are adjacent, then  there is no critical connectivity maintained through their common border, by Lemma~\ref{2sparse}.

Apply Corollary~\ref{rid-blob} to move side-branches in the sparse subcells to the boundary of $M$.
There is only one module that possibly cannot be moved, in the case of two rings that exist in a diagonal configuration and must be connected.
If a new ring is created, we apply Lemma~\ref{3-4rings}.
Now the only branches along interior borders of subcells belong to the two rings, with the possible exception of one module at the middle of $M$.
We can use corner pops and/or staircase moves and Corollary~\ref{rid-blob} to move the interior ring sides to the boundary of $M$ while maintaining connectivity.  This happens regardless of the relative position of the rings or the presence of the extra module.

What remains is to maintain our shortest path requirement, if we still do not have a ring in $M$.
In this case, by Lemma~\ref{enough4ring} we know that there was no initial backbone in $M$.
Thus each connected component  of robot within $M$ ``covers'' at most one corner (in other words there is at least one module gap per side).

Note that the modules in the two subrings alone nearly suffice to create a ring in $M$.  Four modules are missing.  We can remove a strip of width 2 from positions where we wish to have a gap in the boundary of $M$, and use parallel-tunneling to position this material in the current gaps.  Essentially we create a temporary ring of width 2.  Then the remaining material can be moved.
\end{proof}

%%%%%%%%%%%%%%%%%%%%%%%%%%%%%%%%%%%%%%%%%%%%%%%

\begin{lem}
\label{exactly1}
If exactly one subcell $S$ of a merged cell $M$ is a ring, then $M$ can become well-merged using $O(1)$ moves.
\end{lem}
\begin{proof}
Without loss of generality let $S$ be at the bottom-left of $M$.
By Lemma~\ref{2sparse}, in the original robot there was no path from the top border of $M$ leading to either of the bottom subcells.    The same holds for the right border of $M$ and the two left subcells.
Therefore the two interior borders between the three sparse subcells do not preserve any connectivity.
We may use Corollary~\ref{rid-blob} to move branches from those interior borders to the boundary of $M$.  Finally we can do the same for the interior sides of $S$.

%First, blobs in  subcells adjacent to $S$ that touch the ring can be moved to the boundary of $M$. Also, along the other two inner borders, we can ensure that only one side contains  modules (absorb arbitrarily from one side to the other, and remove excess to boundary of $M$).
%Now those inner borders either contain a path from the boundary of $M$ to $S$ or not.  In the latter case any existing branches can be moved to the boundary of $M$ because they serve no purpose. In the former case, we can pop the inner sides of $S$ with the inner border paths, to the boundary of $M$.
We may have to redistribute excess internal material from within $S$.  If $M$ has become a ring, this is easy and has been discussed previously.
Otherwise, we can apply Corollary~\ref{rid-blob} to each full row of the internal ring structure.
This can be required at most eight times before a ring is created.

Our shortest path connectivity requirement is preserved directly, by the fact that the internal borders where not necessary for connectivity.
\end{proof}

%%%%%%%%%%%%%%%%%%%%%%%%%%%%%%%%%%%%%%%%%%%%%%

\begin{lem}
\label{0rings}
If no subcell of a merged cell $M$ is a  ring, then $M$ can become well-merged using $O(1)$ moves.
\end{lem}
\begin{proof}
By Lemma~\ref{2sparse}, we know that in the original robot configuration no path existed
from a side of $M$ to either of the two subcells furthest from it.  Therefore all disjoint subgraphs maintained connectivity between  at most two adjacent  external sides of subcells.
More specifically,
% Furthermore, the adjacent sides could not both be interior subcell borders, because this would imply the existence of a ``long'' path in $M$, which  by the pigeon-hole principle would imply that at least one subcell would have been a ring.
%Therefore the ``inner'' corner of the boundary of each subcell is empty.  In other words, there is no connection between points on the boundary of $M$ that happens to use interior borders of the subcells of $M$.
the first type of allowed path connects points that are separated by a corner of $M$ but are also inside the same subcell.  By induction we assume that these points are already connected along the external boundary of their subcell.
The second type connects points that are on the same border side of $M$ (possibly adjacent subcells).  Again by induction we know that they are already connected along the boundary of $M$.
Therefore our shortest path requirement is preserved.

All that remains is to remove excess material from inner borders of subcells.  This material consists of one or two branches per border, each of which is connected to the boundary of $M$.
%These can be staircased and redistributed with our standard procedures.
%
If any such border $B$ does not connect the boundary of $M$ to the mid-point of $M$ (i.e., to a path on another interior border), then any  module along $B$ can be moved to the boundary of $M$ (with staircasing, and absorbing via parallel tunnel).
Along any inner border we can absorb  modules from one cell's boundary to gaps in its adjacent one. Any remaining material can be sent to the boundary of $M$.

So, without loss of generality,  assume that we have at least one full inner border.  In fact if there is one it can be removed, and if there are only two forming a bend, they too can be eliminated.  The only non-trivial case involves a straight path through the middle of $M$, that may also have one or two straight branches (a $T$-shape or a cross).    Let the straight path be vertical.
If we cannot just cut the path in two and collapse, this means that the path is critical for connectivity.
If the critical connection is between points on opposite sides of $M$, then we know that there must be at least three more side lengths of thin blobs in $M$.  Moreover, because we have no subrings, that initial backbone must zig-zag through the current vertical connection (avoiding to make a backbone in any subring). We might have to argue here that parts of the initial backbone always remain connected.  This would imply that we can simply elevator the vertical connecting path over to the boundary.
\end{proof}

%\begin{thm}
%\label{main}
%If a robot $R$ is constructed with blocks of $32\times32$ atoms, we can transform it to a ring with $O(n\log{n})$ atom operations in $O(\log n)$ parallel steps.
%\end{thm}
\begin{thm}
\label{thm:main_result}
Any source robot $S$ can be reconfigured into any target robot $T$
with $O(n\log{n})$ atom operations in $O(\log n)$ parallel steps, if $S$ and $T$ are constructed with blocks of $32{\times}32$ atoms.
\end{thm}
\begin{proof}
Every cell retains the modules that it initially contained and does not interfere with the configuration of the robot outside the cell, until it is time to merge with its neighbors.
A temporary exception to this occurs during Lemma~\ref{mergeblob}.  Therefore that step should be performed in a way so that no interference occurs (i.e., perform only this operation during one time step).
At every time step, we merge groups of four cells, which by induction are
either rings or sparse.  By Lemmas~\ref{3-4rings}--\ref{0rings},
these four cells merge into a ring or sparse cell.
Thus we construct a ring or sparse cell in $O(\log n)$ parallel time steps.

We show that the total number of operations is $O(n\log n)$.
Each subcell containing $m$ atoms can involve $O(m)$ parallel operations
per time step.
Because there are $O(1)$ time steps per level in the recursion,
and all $m_i$ sum to $n$, the total number of operations per recursion level
is $O(n)$.

Now  consider the bounding box $B$ of $S$.  We construct the smallest square $B_2$ of side length $2^h$ that contains $S$ and has the same lower-left corner as $B$.  Our recursive algorithm takes place within $B_2$.
Now consider the last merge of subcells in our algorithm.  The lower-left subcell $L$ could not have contained $S$, because this would imply that $B_2=L$.   Therefore there must have been a path in $S$ from the left side of $B_2$ leading to the two rightmost subcells (or from bottom to two topmost).
This implies that $S$ will become a ring (not sparse).

Because a ring of specific side length has a unique shape as a function of the number of modules it contains, the resulting ring in $B_2$ serves as a canonical form between $S$ and $T$.
\end{proof}

%%%%%%%%%%%%%%%%%%%%%%%%%%%%%%%%%%%%%%%%%%%%%%%
%%%%%%%%%%%%%%%%%%%%%%%%%%%%%%%%%%%%%%%%%%%%%%%
%%%%%%%%%%%%%%%%%%%%%%%%%%%%%%%%%%%%%%%%%%%%%%%
%%%%%%%%%%%%%%%%%%%%%%%%%%%%%%%%%%%%%%%%%%%%%%%
%%%%%%%%%%%%%%%%%%%%%%%%%%%%%%%%%%%%%%%%%%%%%%%
%%%%%%%%%%%%%%%%%%%%%%%%%%%%%%%%%%%%%%%%%%%%%%%

%\section{Labelled Robots   (optional section)}
%
%\begin{itemize}
%\item Introduce the labelled robots. Applications: some robot atoms have a sensor (like a camera) and it is important that the sensor is at a specific place.
%\item Give lower bound.
%\item Give algorithm for labelled case that matches the lower bound.
%\end{itemize}

\section{Generalization to Three Dimensions}
\label{sec-3D}
\begin{comment}
\begin{itemize}
\item The \textit{boundary} consists of all the $12$ edges of the cube in $3D$.
\item The \textit{near-boundary} consists of all module positions face-connected to the cell's boundary (adjacent and on one of the faces).
\item The other definitions then follow (I think).
\item Our elementary moves suffice. The corner pop changes (we may need to pop along multiple dimensions to get an interior corner to the edges of the cube). $O(1)$ pops seem to suffice to pop one corner.
\item Variations of all the Lemmas / Theorems hold. For the pigeonhole argument to work, we need to cover $48$ sides (4 subcubes of a full cube). This can be done using blocks of size $8k\times 8k\times 8k$ ($k=4$ or $k=2$). Redistribution arguments get a lot more tedious.
\end{itemize}
\end{comment}
%
%
%
%
%
We now discuss how to extend the 2D result to 3D.     The main concepts are similar, so we focus on the few technical differences.

In 3D, our reconfiguration takes place in a cube of side length $2^h$.  We recursively divide space into cubes with side lengths divided by 2.  The \textit{boundary} of the cell consists of all module positions that touch an edge of the cell. As in 2D, a branch that lies entirely in the boundary of a cell is a \textit{side-branch}. The \textit{near-boundary} of a cell consists of all module positions that are face-connected to the boundary.

If we manage to fill the boundary of a cell with modules, it is called a ring. Excess modules
fill the cell from bottom to top, analogous to 2D.  The entire bottom horizontal square layer must be filled in order to progress higher.
A 3D sparse cell has modules only 
%on the boundary (which is the exterior surface of a cell).
on edges (in fact this means that every side-branch can be made to connect to a corner, by a temporary trading argument, just like in 2D, Lemma 3).
%In fact, we impose that every side-branch must touch a cell edge.
When we merge 8 subcells, we use the inductive assumption that they are empty, sparse or rings.

\subsection{Block size and shortest paths}
In 2D, the block size was determined in Lemma~\ref{enough4ring},
where the minimum length of a subcell backbone was compared to the circumference
of a cell. 
Recall that the first part of Lemma~\ref{enough4ring} states that if a subcell contains a backbone then a cell becomes a ring.  
%Furthermore the Lemma describes paths between subcells, indicating that
This is a result of the factor of 8 ratio between the cross-section of 2D blocks and modules.
In other words, a path made of blocks is 8 modules wide.

If we want this statement of Lemma~\ref{enough4ring} to apply in 3D, it suffices to set the multiplicative factor at
24;  If subcells have  side length $c$ (which is also the minimum length of their backbone), then the number of modules in a 3D ring is $24c{-}16$.  Therefore the cross-section of a
backbone should be at least $5{\times}5$ modules.
In other words, we establish that a block should be at least $5{\times}5{\times}5$ modules.\\

A 3D  analog of the second statement of Lemma~\ref{enough4ring} also holds. Paths in the original configuration can only connect adjacent subcell sides 
(if a ring cannot be formed).

Let $P$ be a path beginning on an exterior face $F$ of a subcell, without loss of generality, the front-left face on the top of the cell.  Assume that the entire cell 
does not have enough modules to make a ring.   Then
$P$ can only connect to the other top faces, or to the external vertical faces of its own subcell, or to their neighbors to the left and right. In 
other words, it can connect to any of the \textit{legal faces} defined as the faces touched by one of $F$'s corners. A path to any other external face would contradict our assumption.

%Note that $P$ cannot connect to all of the legal faces simultaneously anyway.  Also, 
Consider a subset $Z$ of the robot within a cell, where every path in this subset has the property described above.  $Z$ can be substituted with a different connected set of modules that  runs along the exterior of the cell and maintains connectivity with every position outside of the cell. 
This can be established with a projection argument.

Similar to 2D, any path of the original robot, within a cell that does not become a ring, can be
replaced with a shortest path along the boundary.

\subsection{Connectivity of rings and sparse cells}
In a 2D cell, a connectivity point was any position on the boundary (since any such position could potentially be connected to an adjacent cell).   Such positions were never tampered with.
However in 3D we do not wish to leave modules in the interior of cell faces.  
We have shown that the internal connectivity of a sparse cell can be maintained by pushing
modules to the boundary.  However we have not addressed the issue of connectivity to adjacent cells.

We now argue that in two adjacent cells ``sharing" face $F$, 
no modules need to be inside $F$  to preserve connectivity.\\

Suppose without loss of generality that the two cells are side-by-side ($F$ is vertical).
If at least one of the cells is a ring, the claim is easy to prove: recall that, by Lemma 1, the paths of a sparse cell can be replaced with paths that only traverse the boundary.
Now notice that the side-branches of the sparse cell $A$ that are in the interior of $F$ can be moved anywhere, since they don't affect connectivity. 
On the one hand if they connect to other modules within $A$, this happens along the surface of $A$, which means through a path that touches the adjacent ring. Thus there is an alternate path from any module in the ring to the remaining modules of $A$.   On the other hand if 
side-branches of $A$ are simply ``islands" in the interior of $F$, then they clearly have no connectivity purpose between the cells and they don't critically connect any modules from the ring cell.  Thus they can just be swept to a corner.

Now suppose that both adjacent cells are sparse.
Note that an initial path from the top of the (union of the) two cells to the bottom
would contain enough modules to make at least one of them a ring (by pigeon-hole, 
as in Lemma 2).
If the cells have side-length $2c$, the path would contain at least $50c$ modules, at least half of which are in one of the cells.  This exceeds the $24c{-}16$ bound given previously.
This means that we can avoid having to shortcut through $F$.
Now, with no connectivity requirement between opposite edges of $F$, we can
essentially look at the union of the two sides of $F$ (one in each cell) and treat it as the 2D case;  all paths can be substituted with new paths along the perimeter of a square. \\

We have established that, when we are to merge 8 subcells, we can assume that all modules
are on the subcell edges.   In fact, in a sparse cell, all paths are shortest paths along the edges.  At this point, we can proceed with a description of how to merge while
preserving connectivity.

\subsection{Module size and merging}
Consider any given branch $b$, made of $8{\times}8{\times}8$ modules.
We could consider $b$ as consisting of four quarters, corresponding to  branches of $4{\times}4{\times}4$ modules.   One such component could be stripped away and moved
about, say in a sparse cell, using 
standard 2D moves (see Lemma~\ref{into-boundary}) to reach any desired position in $O(1)$ time. Clearly it is not crucial for connectivity, since the remaining three quarters are in place.
Thus if we wish to displace $b$ to a new position $c$, where $c$ replaces the connectivity
role of $b$, we can use this {\em doubling trick} to move $b$ in parts.  

We use the term ``doubling" instead of ``quadrupling" because the trick only technically requires double the module size (volume) used in basic-moves.
 However, it is conceptually easier to strip away sub-branches of cross-section $4{\times}4$ (even if this
must be done 4 times), rather than working with something like $6{\times}6{\times}6$ that technically suffices for our argument.\\

Consider 8 subcells about to be merged into cell $C$.
If we use the doubling trick, it is easy to replace the existing
paths in $C$ with new ones that only use the perimeter of $C$.

After this, for each face $F$  of $C$, we push modules
that are in the interior of $F$ towards the edges, while making sure we don't lose
connectivity with the  newly merged cell adjacent to $F$.
This may involve some  trading of modules (between $F$ and its adjacent counterpart),
 as in Lemma~\ref{mergeblob}.\\

The main advantage of the doubling trick is that it allows us to avoid explicitly considering how to move around the modules on the internal
borders of the subcells of $C$.  Note that the underlying idea was also used in 2D,  in the last
paragraph of the proof of Lemma~\ref{2rings}.  However,  recall that in 2D we also had to argue why certain connections along internal borders were
not critical (Lemmas~\ref{exactly1} and~\ref{0rings}).    

 Without the double-module trick, we would probably have to increase the block
size to follow the method used for 2D.  This would involve stronger pigeon-hole arguments.
For instance,
suppose that a subcell has side length $c$.  If it is a ring, it contains at least
$12c{-}16$ modules.   The top four subcells would contain $48c{-}64$ modules.
Now consider an initial path (of blocks) running from the top of the cell to one of the bottom 
subcells. Its length is at least $c$.  If the cross-section of the path is at least $7{\times}7$ modules, then the path must contain more modules than four rings.   Thus the pigeon-hole principle tells us that one of these subcells must end up as
a ring. 
%   So for this 3D analog of Lemma 2, we increase the block size to $7{\times}7{\times}7$ modules.\\
% we could make sure that any path from the top of a cell to one of the bottom  subcells would contain enough modules to make at least one of the top four subcells a ring (by pigeon-hole).  

Then, as in 2D, we can argue that there are no critical connections
along internal borders, whenever fewer than two subcell rings exist.  Note that two subcell rings contain nearly enough modules to make the entire cell a ring. Only 16 modules are missing.
Thus we could repeat the arguments of Lemmas~\ref{2rings},~\ref{exactly1},~\ref{0rings}.
%We calculate that the necessary block size for this approach is $7{\times}7{\times}7$.
\\
\\   
We conclude that by using modules and blocks of  constant size
(either $8{\times}8{\times}8$ and $5{\times}5{\times}5$ respectively, or
 $4{\times}4{\times}4$ and $7{\times}7{\times}7$), we can reconfigure 3D Crystalline robots in $O(\log n)$
parallel steps.

\section{Discussion}
\label{discussion}
The number of atoms in our modules and initial blocks can be reduced.
By using  $2{\times}2$ modules instead of $4{\times}4$, some of our basic operations become relatively complicated.  For example, the staircase move cannot be implemented via sliding, but instead involves a form of parallel tunneling to break off strips that are one module wide, and then using those as carrying tools, etc.     Corner pops  also become particularly unattractive.
Reducing the block size has the result that we can no longer rely only on rings and sparse cells to maintain the connectivity of any  robot.  In 2D, we obtain a small set of orthogonal shortcut trees that must be taken into consideration when merging cells.   We expect that the same holds in 3D, although we have not attempted to compute the necessary set of shortcut types.
% In fact, reducing the block size even further just results in more shortcut trees.    The total number of such shortcut shapes is a small constant.
We conjecture
%(only because we have not rigorously analyzed all cases)
that reconfiguration can take place with minimal modules (2 atoms in each dimension)
and no block restriction.
%A contrary result would be surprising.   In fact, for labeled modules, there seems to be a lower bound argument.

%We have determined that $O(n\log n)$  basic operations suffice to reconfigure any robot.  However, we mention here that
Our algorithm can be implemented in $O(n\log n)$ time.  Each subcell contains a constant number of rectangular components, so determining their relative configuration and series of motions should require constant time.
We also claim that our results extend to the case of labeled robots.
%where specific atoms (or at least modules) must reach particular locations.
This would involve a type of merge-sort using staircase moves, once a straight path of modules is constructed using our algorithm.
%The details remain to be verified.

%Two main open questions remain.
%First, does a similar result hold in 3D?
%Our algorithm does not seem well suited for this case.
We have not determined if %a similar result will hold in 3D, or if
%Second, is
$O(\log n)$
%parallel
steps are worst-case optimal.
Such a lower bound can be given for labeled robots,
by a simple Kolmogorov argument: there exist permutations that contain $\Theta(n \log n)$
bits of information. Each parallel move
can be encoded in $O(n)$ bits (for each robot in order, which sides perform
which operations), so we need $\Omega(\log n)$ steps.

%-------------------------------------------------------------------------
% Bibliography and index
\bibliography{cbots}
\bibliographystyle{abbrv}   %abbrv

\end{document}